\documentclass[%
 reprint,
superscriptaddress,
 amsmath,amssymb,
 aps,
 prl,
]{revtex4-2}

\usepackage{graphicx}
\usepackage{dcolumn}
\usepackage{bm}
\usepackage[colorlinks=true, letterpaper=true, pdfstartview=FitV, linkcolor=blue, citecolor=blue, urlcolor=black]{hyperref}
\usepackage{verbatim}

\usepackage{float}
\usepackage{enumerate}

\usepackage[final]{pdfpages}
\makeatletter
\AtBeginDocument{\let\LS@rot\@undefined}
\makeatother
\begin{document}

\title{Stochasticity-induced non-Hermitian skin criticality }

\author{Xiaoyu Cheng}
\address{College of Physics and Electronic Engineering, Shanxi Normal University, Taiyuan 030031, China}
\address{State Key Laboratory of Quantum Optics Technologies and Devices, Institute of Laser Spectroscopy, Shanxi University, Taiyuan 030006, China}
\address{Science, Mathematics and Technology (SMT) Cluster, Singapore University of Technology and Design, Singapore 487372}
\address{Collaborative Innovation Center of Extreme Optics, Shanxi University, Taiyuan 030006, China}

\author{Hui Jiang}
\email{phyjianghui@dlut.edu.cn}
\address{School of Physics, Dalian University of Technology, Dalian 116024, China}
\address{Department of Physics, National University of Singapore, Singapore 117542}

\author{Jun Chen}
\address{State Key Laboratory of Quantum Optics Technologies and Devices, Institute of Theoretical Physics, Shanxi University, Taiyuan 030006, China}
\address{Collaborative Innovation Center of Extreme Optics, Shanxi University, Taiyuan 030006, China}

\author{Lei Zhang}
\email{zhanglei@sxu.edu.cn}
\address{State Key Laboratory of Quantum Optics Technologies and Devices, Institute of Laser Spectroscopy, Shanxi University, Taiyuan 030006, China}
\address{Collaborative Innovation Center of Extreme Optics, Shanxi University, Taiyuan 030006, China}

\author{Yee Sin Ang}
\email{yeesin\_ang@sutd.edu.sg}
\address{Science, Mathematics and Technology (SMT) Cluster, Singapore University of Technology and Design, Singapore 487372}

\author{Ching Hua Lee}
\email{phylch@nus.edu.sg}
\address{Department of Physics, National University of Singapore, Singapore 117542}

\begin{abstract}
Typically, scaling up the size of a system does not change the shape of its energy spectrum, other than making it denser.
Exceptions, however, occur in the new phenomenon of non-Hermitian skin criticality, where closely competing generalized Brillouin zone (GBZ) solutions for non-Hermitian state accumulation give rise to anomalously scaling complex spectra.
In this work, we discover that such non-Hermitian criticality can generically emerge from stochasticity in the lattice bond orientation, a surprising phenomenon only possible in 2D or beyond. 
Marked by system size-dependent amplification rate, it can be physically traced to the proliferation of feedback loops arising from excess local non-Hermitian skin effect (NHSE) accumulation induced by structural disorder.
While weak disorder weakens the amplification as intuitively anticipated, stronger disorder enigmatically strengthens the amplification almost universally, scaling distinctly from conventional critical system.
By representing cascades of local excess NHSE as ensembles of effectively coupled chains, we analytically derived a critical GBZ that predicts how state amplification scales with the system size and disorder strength, highly consistent with empirical observations. 
Our new mechanism for disordered-facilitated amplification applies generically to structurally perturbed non-Hermitian lattices with broken reciprocity, and would likely find applications in non-Hermitian sensing through various experimentally mature meta-material platforms.

\end{abstract}

\date{\today}

\maketitle

\noindent\emph{Introduction.---}
In a non-Hermitian system, flux can induce intrinsic amplification directionality, forming ``skin" states accumulating against finite system boundaries \cite{lee2016anomalous,kunst2018biorthogonal,song2019non,jin2019bulk,lee2019anatomy,yao2018edge,yao2018non,yokomizo2019non,zhang2020correspondence,jiang2019interplay,okuma2020topological,arouca2020unconventional,helbig2020generalized,okugawa2020second,zhu2020photonic,zhang2022review,shen2022non,lin2023top,okuma2023non,xue2024topologically,yang2025beyond,brighi2024NON,yoshida2024non,wang2025non,longhi2019probing,kawabata2020non,song2019realspace,yang2022designing,wang2024amoeba}. Mathematically, these skin states are encoded as complex-deformed (non-Bloch) momenta that live in the so-called Generalized Brillouin Zone (GBZ) \cite{yao2018edge,yao2018non,yokomizo2019non,lee2019anatomy,zhang2020correspondence,longhi2019probing,kawabata2020non,song2019realspace,jiang2023dimensional,wang2024amoeba}, a highly successful construct in restoring non-Hermitian topological bulk-boundary correspondences.

While this directed amplification is already a well-known phenomenon -- famously branded as the non-Hermitian skin effect (NHSE) \cite{lee2016anomalous,yao2018edge,yao2018non,kunst2018biorthogonal,song2019non,jin2019bulk,lee2019anatomy,yokomizo2019non,jiang2019interplay,okuma2020topological,helbig2020generalized,zhang2020correspondence,okugawa2020second,zhu2020photonic,zhang2022review,shen2022non,lin2023top,okuma2023non,brighi2024NON,yoshida2024non,wang2025non} -- what remains enigmatic are the \emph{critical transitions} between different GBZ solutions \cite{li2020critical,liu2020helical,yokomizo2021scaling,rafi2022CSE,qin2023universal,liu2024non,meng2025generalized,xu2025exciton,rafi2025critical}. Far from just mathematical curiosities, such non-Hermitian skin criticality dramatically manifests through the  unconventional scaling of amplification rate Im$(E)$, due to competing skin localization scales from various participating GBZs. However, rigorous understanding of such critical skin scaling have so far been restricted to simple 1D models with artificially weak couplings \cite{li2020critical,liu2020helical,yokomizo2021scaling,rafi2022CSE,qin2023universal,liu2024non,xu2025exciton,rafi2025critical,qin2025many}, despite the ability for skin competition to occur in almost \emph{all} heterogeneous non-Hermitian systems with broken reciprocity.

In this work, we discover that in 2D or higher, the hallmark of skin criticality -- nontrivial scaling of eigenenergy spectra $E$ and the amplification Im($E$) represents -- occurs prominently in generic multi-orbital NHSE lattices with structural disorder.  It is interesting that such non-Hermitian criticality, usually associated with GBZ transitions in crystalline systems, can be also induced by stochasticity. This is qualitatively distinct from any known paradigm, such field theories based on non-linear sigma models \cite{ketov2013quantum} and mobility edges from Anderson localization \cite{li2009topological,groth2009theory,guo2010topological,Li2020Topological,stutzer2018photonic,meier2018observation,cui2022photonic,chen2024realization,zhang2019topological,zhang2020non,gu2023observation,zhang2021experimental,lin2022observation,chen2017disorder,qu2024topological}. Being exclusive to higher dimensions, it also harbors dissimilar physical origins from random matrix theory \cite{tao2012topics} and stochastic NHSE \cite{longhi2019topological, longhi2021spectral,longhi2025erratic,hao2025interacting,nelson2024nonreciprocity}.

Our stochasticity-induced non-Hermitian skin criticality, marked by system size-dependent amplification rates Im$(E)$, can be physically traced to the emergence of non-reciprocal feedback loops from positional disorder. Instead of weakening the NHSE, as intuitively expected, randomly perturbed atomic positions (in 2D but not 1D) lead to inevitable variations in bond angles and strengths that result in the proliferation of amplification loops. This will be shown to yield optimal scaling amplification ratios at rather strong disorder, as verified through extensive numerics and spatial eigenstate analysis. Finally, we theoretically back our physical mechanism by constructing an effective model that identifies the effects of generic structural disorder with dimensionally-reduced weakly-coupled NHSE subsystems, albeit without requiring the fine-tuning~\footnote{In conventional critical NHSE systems, unconventional scaling is only observed when the subsystem couplings take on weak values within a certain window} needed in known 1D critical NHSE systems \cite{li2020critical,liu2020helical,yokomizo2021scaling,rafi2022CSE,qin2023universal,liu2024non,xu2025exciton,rafi2025critical}.

\begin{figure}
\centering
\includegraphics[width=0.93\linewidth]{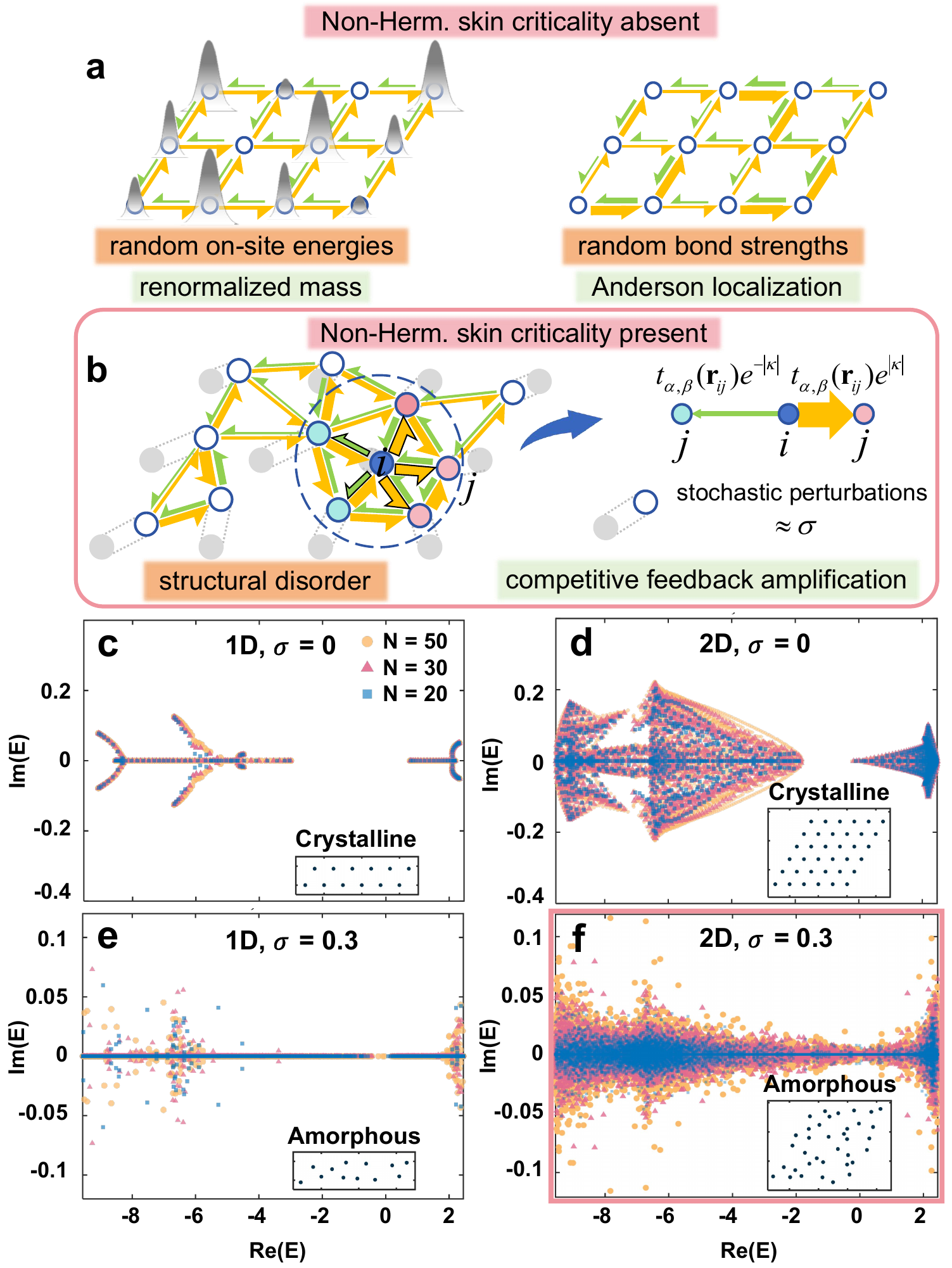}
\caption{
(a) Randon on-site energies or bond strengths lead to phenomena with Hermitian analogs, and \emph{cannot} give rise to non-Hermitian skin criticality by themselves. 
(b) But non-Hermitian criticality can occur due to competitive feedback mechanisms from structural disorder. Randomly perturbing atoms with positional standard deviation $\sigma$ leads to stochastically modified asymmetric left/right inter-atomic hoppings (green/yellow) [Eq.~\ref{HamNH}]. Local inhomogeneities among them yield effective feedback loops. 
(c,d) In both 1D and 2D, the spectrum $E$ of crystalline $H$ [Eq.~\ref{HamNH}] depends very little on the system size $N$, as is typical in most known models. 
(e,f) With structural disorder $\sigma=0.3$, the amplification rate Im$(E)$ in 2D but not 1D increases significantly with $N$ (blue to orange), signaling the prospect of non-Hermitian skin criticality (see discussions surrounding Fig.~\ref{Fig4}). $\kappa=0.15$ for all plots.
}
\label{Fig1}
\end{figure} 

\noindent\emph{Physical motivation for ansatz model with structural disorder.--}
Disorder can manifest in non-Hermitian systems in various ways, such as by randomly perturbing on-site energies, bond strengths or bond hopping non-reciprocity [Fig. \ref{Fig1}(a)]~\cite{zhang2020non,liu2020topological,tang2020topological,mo2022imaginary,liu2021real,kim2021disorder,jiang2019interplay,Non2021Zhang,wanjura2021correspondence}. However, most works have revealed interference or parameter renormalization mechanisms akin to Anderson localization, which are arguably not  unique to non-Hermitian settings. 

This work shall depart from the Anderson localization paradigm and instead focus on long-ranged effects from the interplay of spatial inhomogeneity~\footnote{While smooth spatial hopping inhomogeneity can be captured by a position-dependent GBZ in 1D ~\cite{li2025phase}, the anomalous scaling (skin criticality) in 2D reported in this work precisely invalidates conventional GBZ formulations.} and directed amplification. 
We are primarily concerned with disorder that modulates the direction and distribution of non-Hermitian asymmetric hoppings, such as to give rise to randomly-directed skin states. As rigorously analyzed later before Fig.~\ref{Fig4}, the inevitable competition between these effective NHSE channels is precisely what can cause anomalous scaling dynamics (non-Hermitian skin criticality). Notably, different from the critical NHSE reported in crystalline systems,  \cite{li2020critical,liu2020helical,yokomizo2021scaling,rafi2022CSE,qin2023universal,liu2024non,xu2025exciton,rafi2025critical}, this mechanism does not deliberately involve weak couplings.

\begin{figure*}
\centering
\includegraphics[width=.75\linewidth]{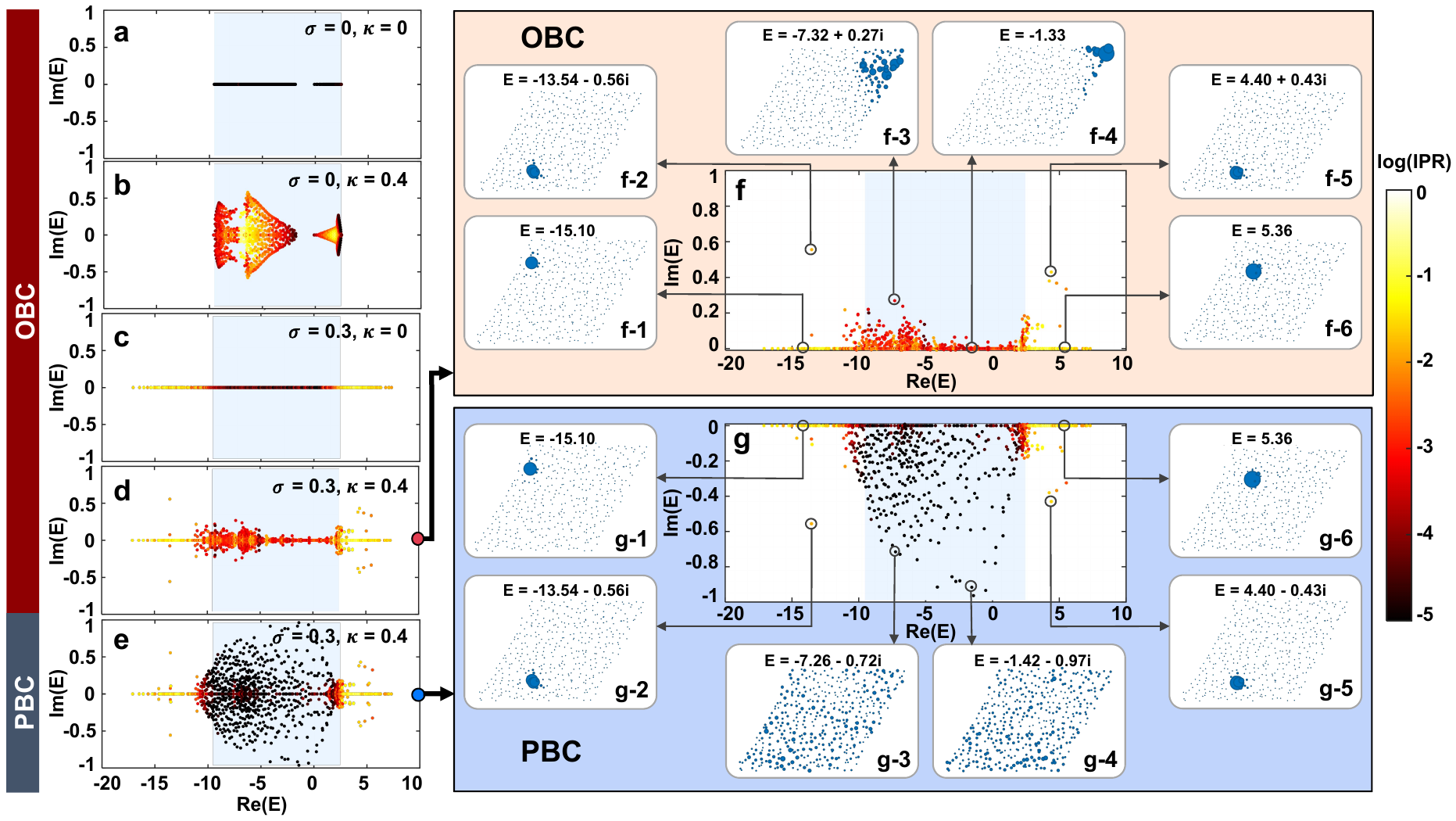}
\caption{Locality and profile of eigenstates under structural disorder. 
All spectra correspond an instance of $H$ [Eq.~\ref{HamNH}] with $N=20$, colored by IPR [Eq.~\ref{EqIPR}]. 
(a) Eigenstates are extended Bloch waves (black) in the crystalline Hermitian case. 
(b) Non-Hermiticity ($\kappa=0.4$) introduces various degrees of skin localization (red to yellow).
(c,d) Disorder ($\sigma=0.3$) introduces Anderson-localized states outside of the original ``skin-dominated" range of Re$(E)$ (pale blue), and also suppresses the amplification rate Im$(E)$ when non-Hermitian.
(e) Switching from OBCs to PBCs makes almost all eigenstates in the skin-dominated range extended (black), but leaves Anderson-localized states untouched.
(f,g) Representative eigenstates for the spectra in (d,e). Saliently, only skin-dominated eigenstates (f-3,f-4) respond to the NHSE and are relevant to non-Hermitian skin criticality.
}
\label{Fig2}
\end{figure*}

A key preliminary insight is that the ingredients for non-Hermitian skin criticality -- asymmetric couplings with randomly perturbed directions and strength -- naturally arise from quenched structural disorder~\cite{duwez1960continuous,zallen2008physics,wang2016quantum,huang2018quantum,wang2022structural,cheng2023topological}, if the couplings are formed from inter-atomic orbital overlap (as in realistic 2D materials \cite{wang2016quantum,wang2012electronics}). 
Since such couplings depend on the relative atomic separation and orientation, displaced atoms would experience random variations in the strengths and directions of their neighboring couplings. 
Our ansatz model takes the form 
\begin{equation}
H= \sum_{i\alpha}\epsilon_{\alpha}c_{i\alpha}^{\dagger}c_{i\alpha}+\sum_{ i\alpha,j\beta}t_{\alpha,\beta}(\mathbf{r}_{ij})e^{\text{sgn}(\mathbf{r}_{ij}\cdot \hat x)|\kappa|}c_{j\beta}^{\dagger}c_{i\alpha},
\label{HamNH}
\end{equation}
where $i,j$ label the atomic sites and $\alpha,\beta$ their atomic orbitals. Each atom is randomly perturbed (isotropically) about its regular lattice position by a distance that is Gaussian-distributed with standard deviation $\sigma$.  
The displacement vector $ \mathbf{r}_{ij}$ between two states $i\alpha$ and $j\beta$ depends only on the relative separation between their atoms $i$ and $j$, which incorporates the stochastic perturbation.  
Importantly, their hopping amplitude consists of the symmetric overlap integral $t_{\alpha,\beta}(\mathbf{r}_{ij})$, which is assumed nonzero between sufficiently near atoms, as well as the non-reciprocal modulation $e^{\text{sgn}(\mathbf{r}_{ij}\cdot \hat x)|\kappa|}$, which takes values of $e^{\mp |\kappa|}$ depending on whether $\mathbf{r}_{ij}$ describes a net leftwards or rightwards displacement [Fig.~\ref{Fig1}b]. $\kappa$ is also known as the hopping asymmetry and controls the directed amplification strength. The onsite orbital energies $\epsilon_{\alpha}$ energetically separate the orbital bands and serve to control how strongly they are intertwined. 

\begin{figure}
\centering
\includegraphics[width=0.87\linewidth]{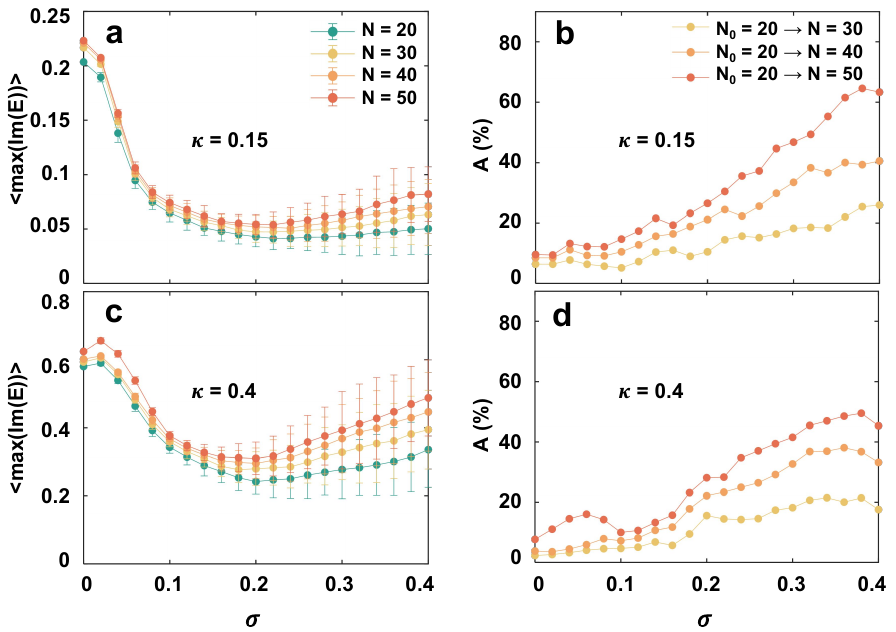}
\caption{Anomalous scaling of amplification rate max(Im$(E)$). (a,c) Averaged max(Im$(E)$) over 400 disorder instances, at different non-Hermiticities $\kappa$. At small disorder $\sigma$, the amplification drops sharply as structural disorder impedes the NHSE. But larger $\sigma$ gives rise to proliferating feedback loops that lead to increasing max(Im$(E)$) with system size $N$. 
(b,d) Corresponding $N$-dependent amplification enhancement ratios $A$ [Eq.~\ref{ANN0}], which can exceed 60\% at large $\sigma$.}
\label{Fig3}
\end{figure}

\begin{figure}
\centering
\includegraphics[width=0.99\linewidth]{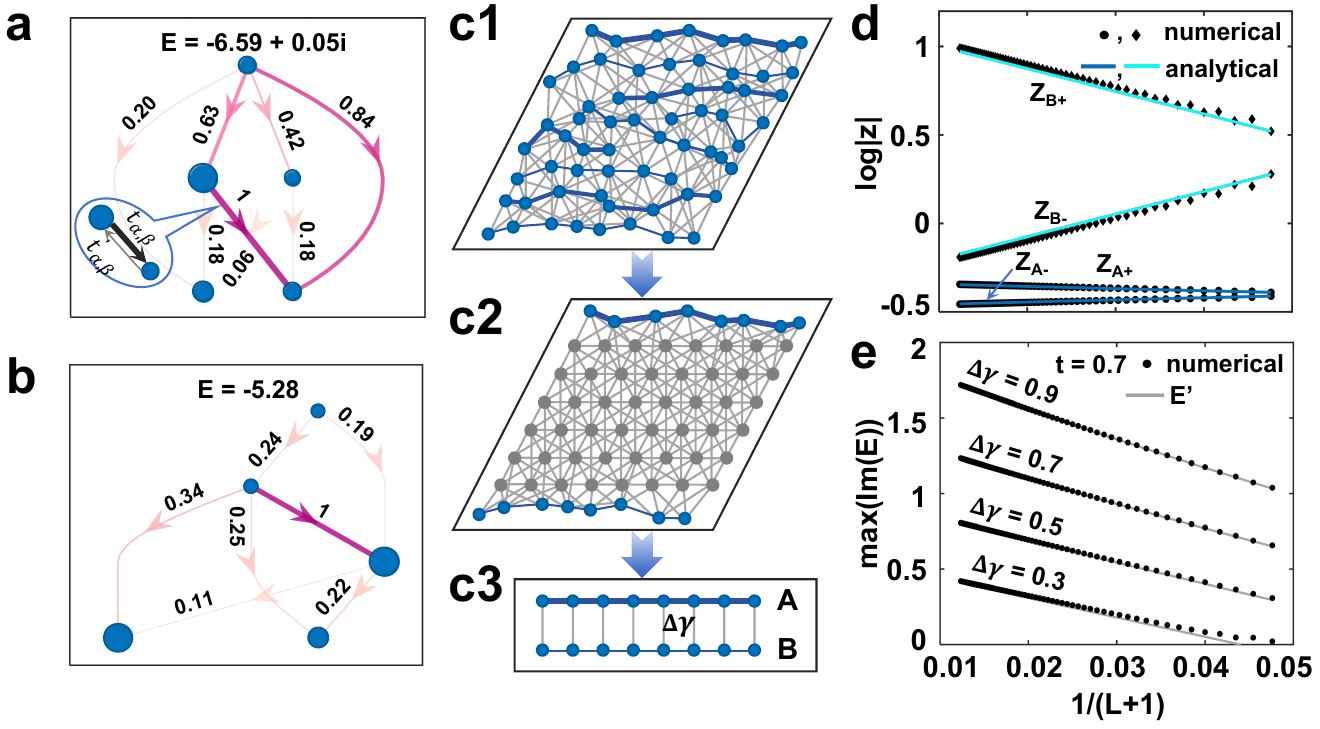}
\caption{Feedback mechanism for stochasticity-induced scaling of amplification.
(a,b) Illustrative eigenstates and their disorder-perturbed hopping weights across orbitals (blue disks) with the largest occupancies. A local feedback loop exists only in (a), where Im$(E)\neq 0$. Larger disks reflect larger state occupancies  $\left|\psi_{i\alpha}\right|^{2}$. Arrows indicate net hopping directions, with color intensity representing effective hopping strength $G = \sqrt{t_{\alpha,\beta} t_{\alpha,\beta}^{'}}$ and line width representing actual transition amplitude $W = G  \sqrt{\left|\psi_{i\alpha} \psi_{j\beta} \right|}$.  
(c1-c3) Cascades of locally asymmetric hoppings form effective NHSE chains. Sufficiently far cascades with nontrivial hopping asymmetry contrast $\Delta \gamma$  can be modeled as a pair of antagonistic weakly-coupled chains $\tilde{\mathcal{H}}$ [Eq.~\ref{EqbA}] exhibiting length $L$-dependent amplification rate. 
(d) $L$-dependent GBZ of $\tilde{\mathcal H}$, with excellent agreement between analytic [Eq.~\ref{Eq4}] and numerical results  from diagonalizing Eq.~\ref{EqbA}. (e) The corresponding amplification rate max(Im$(E)$) increases similarly with $L$ and $\Delta\gamma$, consistent with that in the full stochastic model [Fig.~\ref{Fig3}]. Here, $t=\sqrt{t_{B}/t_{A}}$, $t_{A/B}$ represents the amplitudes of nearest-neighbor hopping to the left/right.
}
\label{Fig4}
\end{figure}

We emphasize that non-Hermitian critical scaling is expected to occur with generically designed $H$ [Eq.~\ref{HamNH}], as long as the orbital structure is sufficiently rich to admit meaningful disorder-induced feedback loops [See Fig.~\ref{Fig4} later]. 
However, for concrete investigation [Figs.~\ref{Fig1} to \ref{Fig3}], we shall specialize to a multi-orbital triangular lattice model with $\alpha\beta \in \{s,p_x,p_y\}$, as detailed in Sec. I of the Supplementary Material (SM) \cite{supp}.

\noindent\emph{Anomalous spectral scaling in higher dimensions.--} 
We first show that structural disorder can give rise to appreciable anomalous spectral scaling, but only in 2D and beyond. This provides initial evidence that distinguishes stochasticity-induced critical NHSE scaling from ``ordinary" critical NHSE scaling in weakly coupled antagonistic 1D chains \cite{li2020critical,liu2020helical,yokomizo2021scaling,rafi2022CSE,qin2023universal,liu2024non,xu2025exciton,rafi2025critical}. Since Im$(E)$ corresponds to the eigenstate amplification rate viz. $|\psi(t)|\sim e^{-\text{Im}(E)t}$, larger Im$(E)$ is symptomatic of more feedback loops. 
Figures.~\ref{Fig1}(c-f) present the complex open boundary condition (OBC) energy spectra for four instances with fixed non-Hermiticity: with or without disorder $\sigma$, and in 1D ($2\times N$ atoms) or 2D ($N\times N$ atoms) configurations.

Crystalline ($\sigma=0$) systems exhibit no qualitative spectral change as $N$ increases from 20 (blue) to 50 (orange), in both 1D and 2D [Figs.~\ref{Fig1}(c,d)].  This is because the regular lattices are designed to have balanced NHSE -- but even if not, there are no weak couplings to break up the lattice into effective feedback loops, as in conventional critical NHSE \cite{li2020critical,liu2020helical,yokomizo2021scaling,rafi2022CSE,qin2023universal,liu2024non,xu2025exciton,rafi2025critical}. 
In 1D, even if structural disorder ($\sigma=0.3$) is introduced [Fig.~\ref{Fig1}(e)], the spectrum merely becomes noisy, without any appreciable scaling of Im$(E)$ with $N$. 

But saliently, structural disorder ($\sigma=0.3$) in the 2D case [Fig.~\ref{Fig1}(f)] causes Im$(E)$ to increase with system size $N$ (blue$\rightarrow$pink$\rightarrow$orange). This suggests that larger amorphous lattices contain stronger feedback loops, which thrive only when there is sufficient ``space" (in 2D but not 1D), despite the accompanying destructive interference that also suppresses amplification (Im$E$).

\noindent\emph{Anatomy of non-Hermitian spectra.--} 
To pinpoint the origin of the Im$(E)$ scaling in our 2D structurally perturbed lattice, we examine its eigenstates in Fig.~\ref{Fig2}. The locality of each eigenstate $\psi_\mu(\bold x)$ is captured by its inverse participation ratio (IPR)~\cite{evers2008anderson,martinez2018non,wang2019non,zhang2020non,longhi2019topological,jiang2019interplay,tang2020topological,mo2022imaginary}
\begin{equation}
\mathrm{IPR}_\mu=\sum_{\bold x}\left|\psi_\mu(\bold x)\right|^{4}/\left({\sum_{\bold x}\left|\psi_\mu(\bold x)\right|^{2}}\right)^{2},
\label{EqIPR}
\end{equation}
as indicated by the colorbar. Very low IPR (dark red or black) indicates an extended state, while very high IPR ($\log\mathrm{IPR}_\mu\approx 0$, yellow) suggests a localized state. For reference, we first show crystalline ($\sigma=0$) OBC scenarios in the absence/presence of the NHSE [Figs.~\ref{Fig2}(a,b): Non-Hermiticity not only makes the spectrum complex, but also partially localizes most eigenstates (orange and red) near the boundary.

Structural disorder ($\sigma=0.3$) introduces additional strongly localized eigenstates (yellow) outside of the original range of Re$(E)$, regardless of non-Hermiticity $\kappa$ or boundary conditions (OBC/PBC), as shown in [Figs.~\ref{Fig2}(c-e)] for the same disorder instance. From Figs.~\ref{Fig2}(f,g), which showcase sample eigenstates from the OBC/PBC spectra of Figs.~\ref{Fig2}(d,e), these yellow eigenstates are ostensibly Anderson-localized, since they are clustered around specific disordered features, and are impervious to the presence of boundaries, or NHSE pumping. 

As such, the most interesting disordered eigenstates lie in the so-called ``skin-dominated region" (blue-shaded). As expected of NHSE states, they are pumped by directed amplification and accumulate towards the right corner under OBCs [Figs.~\ref{Fig2}(f3,f4)]. And under PBCs i.e. without spatial boundaries, they remain as manifestly extended states [Figs.~\ref{Fig2}(g3,g4)]. In this work, we will only focus on the amplification rate Im$(E)$ of these skin-dominated disordered states, since they are precisely the manifestations of stochasticity that are independent of Anderson localization. 

\noindent\emph{Stochasticity-induced amplification from scaling.--} Having identified the ``skin-dominated" eigenstates that stem from non-local directed amplification, we next empirically examine how their amplification is non-monotonically affected by structural disorder. The long-time amplification rate of a generic initial state is given by max(Im$(E)$) across all eigenstates. As intuitively expected from the disruption to directed amplification, max(Im$(E)$) initially decreases as structural disorder $\sigma$ is introduced, as shown for both $\kappa=0.15$ and $0.4$ in Figs.~\ref{Fig3}(a,c), calculated from 400 random instances.

However, beyond $\sigma\approx 0.2$, max(Im$(E)$) can enigmatically start to increase again, particularly for larger system sizes $N$ and non-Hermiticity $\kappa$. This trend is observed very consistently, despite the larger error bars associated with greater structural variations. In particular, max(Im$(E)$) is consistently enhanced by larger $N$, indicative of anomalous critical scaling. This is quantified through the amplification enhancement ratio
\begin{equation}
\label{ANN0}
A
=\frac{\text{max(Im($E$)})_{N}-\text{max(Im($E$)})_{N_{0}}}{\text{max(Im($E$)})_{N_{0}}},
\end{equation}
which expresses the enhancement in amplification rate max(Im$(E)$) at size $N$, compared to that at a reference $N_0$ (See Sec. IV of \cite{supp} for further measures of the localization and amplification). Shown in Figs.~\ref{Fig3}(b,d) are the enhancement ratios $A$ for $N=30,40,50$, relative to $N_0=20$. In the crystalline limit ($\sigma=0$), $A$ almost vanishes, as expected for ordinary lattices. But in general, $A$ is seen to increase with $\sigma$, establishing that non-Hermitian structural disorder indeed leads to increased amplification. Saliently, comparing between $N=20$ and $50$ (red), $A$ can exceed 60\% at strong disorder $\sigma=0.4$.

\noindent\emph{Emergent amplificative feedback from stochasticity.--} We now suggest the physical mechanism behind the anomalous max(Im$(E)$) scaling with system size $N$. In the crystalline limit with discrete translation symmetry, the NHSE acts uniformly and can be simply encoded as a basis change \cite{yao2018edge,yao2018non,yokomizo2019non,lee2019anatomy,zhang2020correspondence,longhi2019probing,kawabata2020non,song2019realspace,jiang2023dimensional} \footnote{More sophisticated GBZ deformations \cite{li2025phase} have also been developed to encode multi-orbital or even spatially inhomogeneous NHSE, but these approaches are so far limited to 1D.}. However, structural disorder 
make certain bond couplings become stronger or more asymmetric (non-reciprocal) than their neighbors. Bond configurations containing net non-reciprocal closed cycles become local feedback loops that lead to amplification, i.e., $\text{Im}(E)>0$, as shown in Fig.~\ref{Fig4}(a) vs. (b) with illustrative eigenstates, and further showcased in Sec. V of the SM \cite{supp}.

To relate this feedback mechanism with anomalous critical scaling of amplification, a key insight is to recognize that, at sufficiently large disorder density (i.e. $\sigma \gtrapprox0.2$), cascades of excess local hopping asymmetries join to form effective NHSE chains, with their mean length $L$ scaling proportionally with $N$ [Fig.~\ref{Fig4}(c1)]. 
Each pair of chains $A$ and $B$ that are reasonably aligned [Figs.~\ref{Fig4}(c2-c3)] can be modeled as
\begin{equation}\label{EqbA}
 \begin{split}
&\qquad\tilde{\mathcal H}= H_A\oplus H_B +\sigma_x\otimes \tilde{\delta},
\end{split}
\end{equation}
where $H_\eta=\sum_x t_{\eta}\left[e^{\gamma_{\eta}}|x\rangle \langle x+1|+e^{-\gamma_{\eta}}|x\rangle \langle x-1|\right]$, with $t_{\eta}\exp(\pm \gamma_{\eta}) $ representing the amplitudes of left/right hoppings in chain $\eta=A, B$. In this stochastic backdrop, the hopping asymmetries $\gamma_\eta$ and amplitudes $t_\eta$ take on a wide range of values, but the effective inter-chain coupling $\tilde \delta$ is generically weak unless the chains are adjacent \cite{qin2024geometry}.
Note that in this description, we have already ``gauged-out" spatial inhomogeneities in the hopping asymmetry, such that $\gamma_\eta$ can be taken as constant across the chain. However, the generically different amplitudes $t_\eta$ prevent $\tilde{\mathcal H}$ from being reduced to well-studied critical NHSE models \cite{li2020critical}. 

Careful mathematical analysis of the boundary matrix (see Sec. VI of the SM \cite{supp}) reveals that the dominant (max(Im$(E)$)) OBC eigenfunctions of $\tilde{\mathcal H}$ take the form $\psi_{\eta}(x)\sim \sum_\pm c_{\eta\pm} z_{\eta\pm}^x$, where $ c_{\eta\pm}$ are constants determined by the boundary conditions, and 
\begin{footnotesize}
\begin{equation}
\begin{aligned}\label{Eq4}
\log|z_{A\pm}|&=-\Delta\gamma \pm \left(\text{Im}(F)+\text{Im }\left(\text{atan}\left(\frac{1-t^2}{1+t^2} \cot (F) \right)\right)+i\frac{\pi}{2}\right),\\
\log|z_{B\pm}| &= \Delta\gamma \pm \left(\text{Im}(F)-\text{Im }\left(\text{atan}\left(\frac{1-t^2}{1+t^2} \cot (F) \right)\right)+i\frac{\pi}{2}\right),\\
\end{aligned}
\end{equation}
\end{footnotesize}
where $\Delta\gamma=\frac{\gamma_{A}-\gamma_{B}}{2}$, $t = \sqrt{t_{B}/t_{A}}$ and $iF=i\pi/2 +\gamma+\frac{1}{L+1}\log\left(-2i\delta \text{e}^{-2\gamma}\right)$.
Interestingly, there is not one, but two distinct effective ``momenta" i.e., GBZ solutions $-i\log z_{\eta\pm}$ (instead of a single conventional GBZ), a consequence of the irremovable disorder-induced NHSE competition between the two chains. This antagonism also underscores the universal existence of the $1/(L+1)$ factor in $z_{\eta\pm}$, which appears regardless of the effective parameters $t_\eta$ and $\gamma_\eta$. The accuracy of Eq.~\ref{Eq4} is verified through its close fit to numerical results in Fig.~\ref{Fig4}(d), where the system size dependence is apparent. By substituting these analytical and numerical GBZ data into Eq.~\ref{EqbA}, we also obtain amplification rates max(Im$(E)$) that clearly increase with $L$ and $\Delta\gamma$. These trends are completely consistent with the empirical results in Figs.~\ref{Fig3}(a,c), at least for $\sigma>0.2$, since the hopping asymmetry contrast $\Delta\gamma$ increases with disorder $\sigma$ and hopping asymmetry $\kappa$.

\noindent\emph{Discussion.--} For the first time, structural disorder is shown to produce a new amplification mechanism that is prominently sensitive to system size. It hinges on the subtle competition between perturbed non-Hermitian feedback loops, which are absent in previous studies of non-Hermitian disorder \cite{zhang2020non,liu2020topological,tang2020topological,mo2022imaginary,liu2021real,kim2021disorder,jiang2019interplay,Non2021Zhang}. 
By representing excess antagonism of NHSE cascades with effective coupled-chains, we analytically reproduced the system-size dependence of the GBZ and Im$(E)$. The derived GBZ exists as a critically coupled solution pair, and rise only from a parent amorphous system in 2D or higher. 

Importantly, our stochastic-induced NHSE criticality mechanism is not restricted to the ansatz model. 
Being fundamentally based on the emergent competition between otherwise orthogonal NHSE paths, it should in general arise from the breaking of orientational or nematic order in generic non-Hermitian systems with broken reciprocity. Marked by disorder-facilitated amplification, we expect it to drive a new avenue of non-Hermitian sensing in photonic \cite{longhi2015robust,feng2017non,zhu2020photonic,parto2020non,li2023exceptional,lin2024observation,ye2025observing,wang2025nonlinear}, acoustic \cite{gu2021controlling,zhang2023observation,huang2024acoustic,gu2022transient,wu2025hybrid,hu2025acoustic}, quantum circuit \cite{wang2024absence,shen2025observation,liu2024dynamical,koh2025interacting} 
and electrical \cite{hofmann2020reciprocal,helbig2020generalized,zou2021observation,su2023simulation,rafi2025frequency,liu2021non,yuan2023non,zhou2025observation} platforms, all of which have seen much recent success in emulating non-Hermitian condensed matter due to their versatility and controllability.

This work was supported by the National Natural Science Foundation of China (Grant No. 12474047, 12174231, 12074230), the Fund for Shanxi ``1331 Project", Fundamental Research Program of Shanxi Province through 202103021222001 and the Graduate Education Innovation Project of Shanxi Province (2023KY013). X.C. was supported by the China Scholarship Council (CSC). This research was partially conducted using the High Performance Computer of Shanxi University. C.H.L acknowledges support from the Ministry of Education, Singapore (MOE Tier-II award numbers: MOE-T2EP50222-0003 and MOE-T2EP50224-0021) and (Tier-I WBS number: A-8002656-00-00). Y.S.A. acknowledges the support from the Ministry of Education, Singapore (Award No.: MOE-T2EP50224-0021).

\bibliography{ref}

\clearpage


\section*{Supplementary material}

# Supplemental Materials for "Stochasticity-induced non-Hermitian skin criticality"

Xiaoyu Cheng,[1, 2, 3, 4] Hui Jiang,[5, *] Jun Chen,[6, 4] Lei Zhang,[2, 4, †] Yee Sin Ang,[3, ‡] and Ching Hua Lee[5, §]

[1]*College of Physics and Electronic Engineering, Shanxi Normal University, Taiyuan 030031, China*
[2]*State Key Laboratory of Quantum Optics and Quantum Optics Devices,*
*Institute of Laser Spectroscopy, Shanxi University, Taiyuan 030006, China*
[3]*Science, Mathematics and Technology (SMT) Cluster,*
*Singapore University of Technology and Design, Singapore 487372*
[4]*Collaborative Innovation Center of Extreme Optics, Shanxi University, Taiyuan 030006, China*
[5]*Department of Physics, National University of Singapore, Singapore 117542*
[6]*State Key Laboratory of Quantum Optics and Quantum Optics Devices,*
*Institute of Theoretical Physics, Shanxi University, Taiyuan 030006, China*

This supplementary material is arranged by sections as follows:

## I. Designing a model that exhibits non-Hermitian skin criticality

This section introduces the non-Hermitian lattice model with stochastic bond orientation as the system for study. It establishes the key ingredients of structural disorder and broken reciprocity necessary to induce the novel critical scaling phenomena.

## II. Necessary conditions for nontrivial spectral scaling

This section compares 1D and 2D non-Hermitian systems, revealing the dimension-dependent nature of stochasticity-induced non-Hermitian skin criticality: in 2D systems, the average maximum imaginary part significantly increases with system size, foreshadowing non-Hermitian critical scaling, whereas no clear size dependence is observed in 1D systems.

## III. Behavior changes in parameter space

This section systematically investigates how the system's energy spectrum and eigenstate distribution evolve across the key parameters—non-reciprocal coupling strength ($\kappa$), structural disorder strength ($\sigma$), and system size ($N$)—revealing an anomalous spectral scaling behavior.

## IV. Comparison of the physical properties in skin-dominated and full energy regions

This section calculates two additional physical quantities in the skin-dominated region, explains the unique behavior changes of the anomalous scalar, and compares the skin-dominated region with the full energy region, showing that the anomalous critical scalar lies in the skin-dominated region and is independent of the Anderson localization.

## V. Understanding feedback loops in real space

This section physically identifies the real-space origin of skin criticality by revealing how structural disorder generates local closed feedback loops with net non-reciprocal circulation.

## VI. Analytical explanation of skin competition

An analytical theory is developed by mapping the disordered system to an ensemble of effectively coupled chains, successfully deriving a critical Generalized Brillouin Zone (GBZ). This critical GBZ accurately predicts how state amplification scales with both system size and disorder strength, providing a fundamental explanation for the observed numerical results.

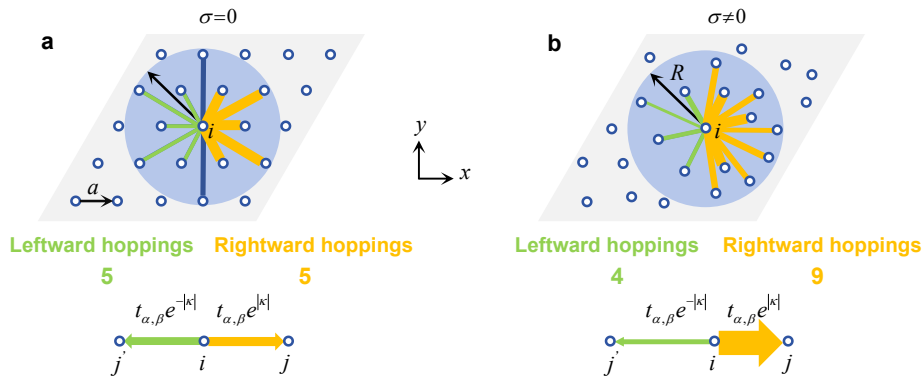

FIG. S1. Schematic diagram of crystalline and amorphous systems. (a) Schematic diagram of the triangular lattice model consisting of $5 \times 5$ lattice points (crystalline structure, $\sigma = 0$). The shaded area represents the coupling region with cutoff radius $R$ centered at atom $i$. The yellow lines represent hopping from atom $i$ to the right neighbor $j$, denoted as $t_{\alpha,\beta} e^{|\kappa|}$; the green lines represent hopping to the left neighbor $j'$, denoted as $t_{\alpha,\beta} e^{-|\kappa|}$, and the blue lines represent couplings perpendicular to the x-axis, denoted as $t_{\alpha,\beta}$. (b) Schematic diagram of an amorphous structure with random atomic displacement ($\sigma \neq 0$). The number of leftward and rightward hoppings from the atom $i$ is statistically analyzed. When $\sigma = 0$, the numbers of leftward and rightward hoppings are equal. In contrast, when $\sigma \neq 0$, the presence of random disorder breaks the balance, leading to an asymmetric distribution of hoppings. The width of the arrows indicates the magnitude of the coupling count. Here, the orbital degrees of freedom are simplified for clarity.

# I. DESIGNING A MODEL THAT EXHIBITS NON-HERMITIAN SKIN CRITICALITY

## Quick survey of non-Hermitian disorder in existing works

In non-Hermitian systems, disorder can manifest in various forms. Much of existing literature focuses on random on-site disorder, which can induce topological phase transitions [1–3]. For example, Kim $et\ al.$ [1] showed that random disorder term breaks the translational symmetry and induces distinctive energy-dependent phase transition by renormalization of the topological mass term. Furthermore, nonreciprocal hopping, which arises due to time-reversal symmetry breaking, can lead to effects such as unidirectional transport [4] and topological phenomena [5, 6]. Tang $et\ al.$ [5] revealed that the nonreciprocal hopping can enlarge the topological regions induced by disorders. Ref [6] also revealed that the combination of disorder and nonreciprocal hopping can induce non-Hermitian topological Andersson insulators that preserve the chiral symmetry, and also the well-known Anderson localisation phenomenon occurs [7, 8], etc. However, most of the work is based on periodic lattice systems with nearest-neighbor coupling or is protected by certain symmetries.

## Our model

In two dimensions or higher, the non-Hermitian critical scaling caused by the stochasticity in lattice bond orientations that we have proposed is generally applicable to non-Hermitian lattices with structural perturbations that break reciprocity, and are not limited to specific models. For concreteness, we shall use the following multi-orbital triangular lattice model inspired by real materials, such as Au/GaAs[111] [9]:

$$H = \sum_{i\alpha} \epsilon_\alpha c_{i\alpha}^\dagger c_{i\alpha} + \sum_{i\alpha, j\beta} t_{\alpha,\beta}(\mathbf{r}_{ij}) e^{\mathrm{sgn}(\mathbf{r}_{ij} \cdot \hat{x})|\kappa|} c_{j\beta}^\dagger c_{i\alpha}, \tag{1}$$

where $i, j$ label the atomic sites and $\alpha, \beta$ their atomic orbitals (each atom has three orbitals $(s, p_x, p_y)$), $c_{i\alpha}^\dagger (c_{i\alpha})$ represents the creation (annihilation) operators at site $i$ with $\alpha$ orbital. The first term $\epsilon_\alpha$ describes the on-site energy of $\alpha$ orbital, energetically separates the orbital bands and serves to control how strongly they are intertwined. The second term $t_{\alpha,\beta}(\mathbf{r}_{ij})$ represents the hopping amplitude between two atoms $i\alpha$ and $j\beta$, calculated from the symmetric overlap integral, which is given by

$$t_{\alpha,\beta}(\mathbf{r}_{ij}) = \frac{\Theta(R - r_{ij})}{r_{ij}^2} \mathrm{SK}\left[V_{\alpha\beta\delta}, \hat{\mathbf{r}}_{ij}\right], \tag{2}$$

where $R$ represents the cutoff radius (set arbitrarily at $R = 1.9a$). There is explicit dependence on the orbital type ($\alpha$ and $\beta$) and the intersite vector $\mathbf{r}_{ij}$. For simplicity, we work in units where the lattice constant $a$ is equal to 1. The atom $i$ interacts with atoms within a radius $R$ centered around it, including both nearest-neighbor interactions and long-range interactions, as shown in the blue shaded region in Fig. S1(a). Here, we assume that the coupling between atoms is non-zero only if they are sufficiently close, as implemented by $\Theta(R - r_{ij})$, the step function. SK[·] represents Slater-Koster parametrization [10] for three orbitals, which is customarily used in solid-state physics to describe the electronic band structure in tight-binding models. The matrix elements of SK parameterization take the form

$$\text{SK}\left[V_{\alpha\beta\delta}, \hat{\mathbf{r}}_{ij}\right] =$$
$$\begin{pmatrix} V_{ss\sigma} & sin\varphi cos\theta \; V_{sp\sigma} & sin\varphi sin\theta V_{sp\sigma} \\ -sin\varphi cos\theta V_{sp\sigma} & (sin\varphi cos\theta)^2 V_{pp\sigma} + [1 - (sin\varphi cos\theta)^2]V_{pp\pi} & (sin\varphi)^2 cos\theta sin\theta(V_{pp\sigma} - V_{pp\pi}) \\ -sin\varphi sin\theta V_{sp\sigma} & (sin\varphi)^2 cos\theta sin\theta(V_{pp\sigma} - V_{pp\pi}) & (sin\varphi sin\theta)^2 V_{pp\sigma} + [1 - (sin\varphi sin\theta)^2]V_{pp\pi} \end{pmatrix},$$
$$(3)$$

Each bond integral $V_{\alpha\beta\delta}$ primarily depends on the interatomic distance, where $\varphi$ is the polar angle measured from the positive end of the z-axis; $\theta$ is the azimuthal angle measured from the x-axis in the counterclockwise direction. $\hat{\mathbf{r}}_{ij}$ is the unit direction vector. We use the SK parameterization method to give the on-site energies of each orbital $\epsilon_s = 1.8$ eV, $\epsilon_p = -6.5$ eV, the other parameters of couplings between orbits at different positions are $V_{ss\sigma} = -0.256$ eV, $V_{sp\sigma} = 0.576$ eV, $V_{pp\sigma} = 1.152$ eV, $V_{pp\pi} = 0.032$ eV.

Of key importance is the non-reciprocal modulation factor in the second term, $e^{\text{sgn}(\mathbf{r}_{ij} \cdot \hat{x})|\kappa|}$, which takes values of $e^{\mp|\kappa|}$ depending on whether $\mathbf{r}_{ij}$ describes a net leftwards or rightwards displacement. If the x-coordinate of atom $j$ is to the right (left) of the coordinate of atom $i$, then $\mathbf{r}_{ij}$ has a net rightwards (leftwards) displacement, and the corresponding coupling modulation factor takes a positive (negative) value, $t_{\alpha,\beta}e^{|\kappa|}$ ($t_{\alpha,\beta}e^{-|\kappa|}$), as shown by the yellow (green) lines in Fig. S1(a). When atoms $i$ and $j$ have the same x-coordinate, the coupling term between them is $t_{\alpha,\beta}$, as shown by the blue lines in Fig. S1(a). $\kappa$ is also known as the hopping asymmetry, and controls the directed amplification strength. Below, the number of hoppings for atom $i$ leftward and rightward is counted. When $\sigma = 0$, the number of hoppings leftward and rightward is equal, indicating a uniform distribution of hoppings in the system.

To generate a quenched configuration with spatial disorder, assume that each atom is randomly perturbed (isotropically) about its regular lattice position with the displacement following a Gaussian distribution with a standard deviation of $\sigma$. The movement of atomic positions is represented by vectors $\mathbf{r}_{ij}$, which depend only on the relative separation between atoms $i$ and $j$. Stochastic perturbations of atomic positions cause randomness in the system, which inevitably leads to variations in the bond angles and bond strengths between atoms, as shown in Fig. S1(b). In this illustrative example, after a random perturbation, the number of atoms hopping to the left and right side of atom $i$ changes to 4 and 9, respectively. In comparison to the crystalline system, the number of couplings on both the left and right sides remains at 5. This clearly demonstrates that randomness affects the number of atoms coupled with atom $i$, resulting in an unbalanced coupling distribution within the system.

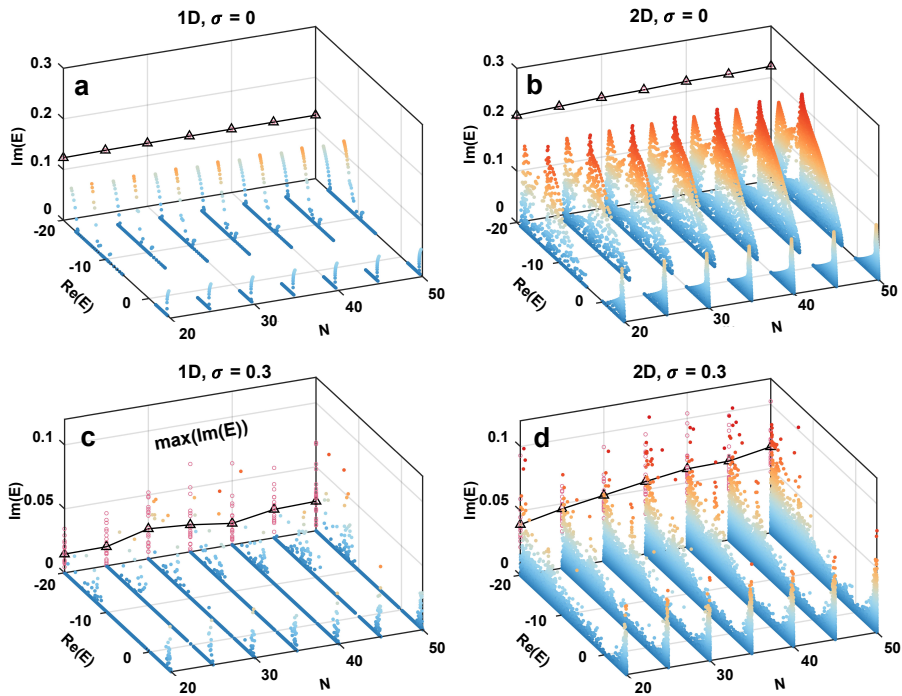

FIG. S2. Dimension-dependent nature of stochasticity-induced non-Hermitian skin criticality. Energy spectra of Eq. 1 in 1D (a) and 2D (b) systems as a function of size $N$ when $\kappa = 0.15$ and $\sigma = 0$. (only the upper half of the spectrum is shown). Energy spectra of 1D (c) and 2D (d) systems as a function of size $N$ when $\kappa = 0.15$ and $\sigma = 0.3$. The color coding represents the size of the imaginary part. The redder the color, the larger the imaginary part. The pink circles projected on the back wall are the projections of the eigenstate with the largest imaginary part in the skin-dominated region of the energy spectrum, and the black triangles represent the average value of the projection points. In (c) and (d), the energy spectra of 20 different structural configurations are accumulated at each size.

## II. NECESSARY CONDITIONS FOR NONTRIVIAL SPECTRAL SCALING

In Fig. S2, we compare the spectra of our system in 1D and in 2D, with and without structural disorder ($\sigma$). Saliently, only the 2D case (d) with $\sigma \neq 0$ exhibits marked changes in the spectrum as system size $N$ increases. Here we only focus on the skin-dominated eigenstates, and combined the spectral points from 20 different disorder instances in each spectral plot slice.

In our color scheme, the pink circles indicate the eigenstates with the maximum imaginary in the skin-dominated region of the energy spectrum, and the black triangle represents the averaged max(Im(E)). When $\sigma = 0$, as shown in Figs. S2(a,b), max(Im(E)) is almost unaffected as the size $N$ increases (whether 1D or 2D). However, when disorder is taken into account, for the 1D system [Fig. S2(c)], the average value of the maximum value of the imaginary part (black curve) has no obvious dependence on the size, while for the 2D system [Fig. S2(d)], the average value of the maximum value of the imaginary part shows a significant dependence on the system size $N$. As the size increases, the imaginary part also increases, indicative of non-Hermitian critical scaling. This is because the increase in dimensionality and size provides sufficient "space" for states to thrive and leads to stronger feedback loops, although the presence of random disorder also suppresses the amplification of the imaginary part (as elaborated in other sections).

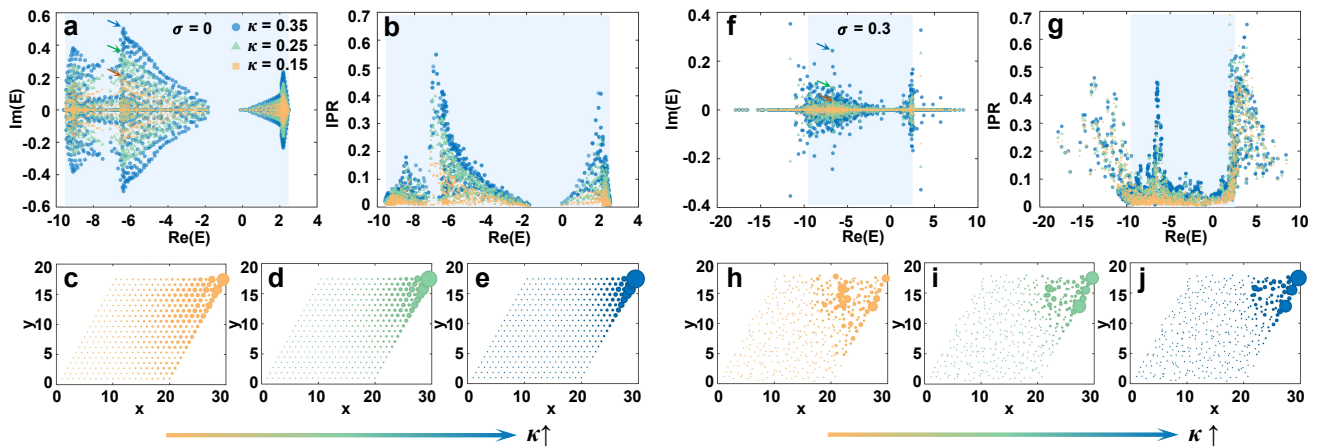

FIG. S3. The influence of non-reciprocal coupling strength $\kappa$ on the energy spectrum. (a) Energy spectra of Eq. 1 at different $\kappa$ when $\sigma = 0, N = 20$. (b) Corresponding IPR [Eq. 2 in the main text] at different $\kappa$. (c)-(e) The state density distribution of the eigenstate pointed by the arrow in (a). (f)-(j) Similar to (a)-(e), except $\sigma = 0.3$.

## III. BEHAVIOR CHANGES IN PARAMETER SPACE

The parameter space has three main adjustable variables: the non-reciprocal coupling strength $\kappa$, the structural disorder strength $\sigma$, and the system size $N$. Figures S3, S4, and S5 show the effects of $\kappa$, $\sigma$, and $N$ on the system energy spectrum and state distribution.

In Fig. S3, when $\sigma = 0$, [Figs. S3(a-e)], for the same size ($N = 20$), as $\kappa$ increases, the width of the energy spectrum on the imaginary axis increases, and the wave function is directionally amplified and pumped to the corners of the system, exhibiting a stronger NHSE. When disorder is introduced, [Figs. S3(f-j)], based on the same disordered structural configuration, as $\kappa$ increases, the maximum value of the imaginary part increases on the imaginary axis, and the IPR of the wave function increases. In other words, regardless of whether there is structural disorder or not, when the non-reciprocal coupling strength is stronger, both the energy spectrum and the wave function will exhibit a stronger "skin effect".

In Fig. S4, with fixed non-reciprocal strength and size, as $\sigma$ increases, more strongly localized eigenstates appear at both ends of the real axis of the energy spectrum, which have higher IPRs. Furthermore, from the spectra, in the skin-dominated region, the maximum value of the imaginary part shows a non-monotonic change as the disorder strength increases.

In Fig. S5, when the non-reciprocal strength is fixed and $\sigma = 0$, [Figs. S5(a-e)], as the size increases, the energy spectrum and wave function distribution have no qualitative spectral change. But saliently, when disorder exists (at sufficiently large disorder density, $\sigma = 0.3$), [Figs. S5(f-j)], the maximum value of the energy spectrum increases significantly with increasing size. It can also be seen that the wave function is pumped to more corner areas at large sizes, which is a phenomenon that does not occur in crystalline systems. While stronger non-reciprocity enhances the skin effect, the crucial phenomenon of size-dependent amplification emerges only under sufficient disorder, where increased system size significantly boosts the maximum imaginary energy due to proliferated feedback loops—a hallmark of skin criticality unique to 2D systems.

In summary, as $\kappa$ increases, the directional amplified pump eigenstates become more localized. The increased size leads to a surge in feedback loops, and the presence of disorder leads to a diversity of structural configurations, making the system no longer uniform in pump intensity. These different behaviors in parameter space also hold the potential for discovering critical scaling in 2D systems that differs from that in 1D systems.

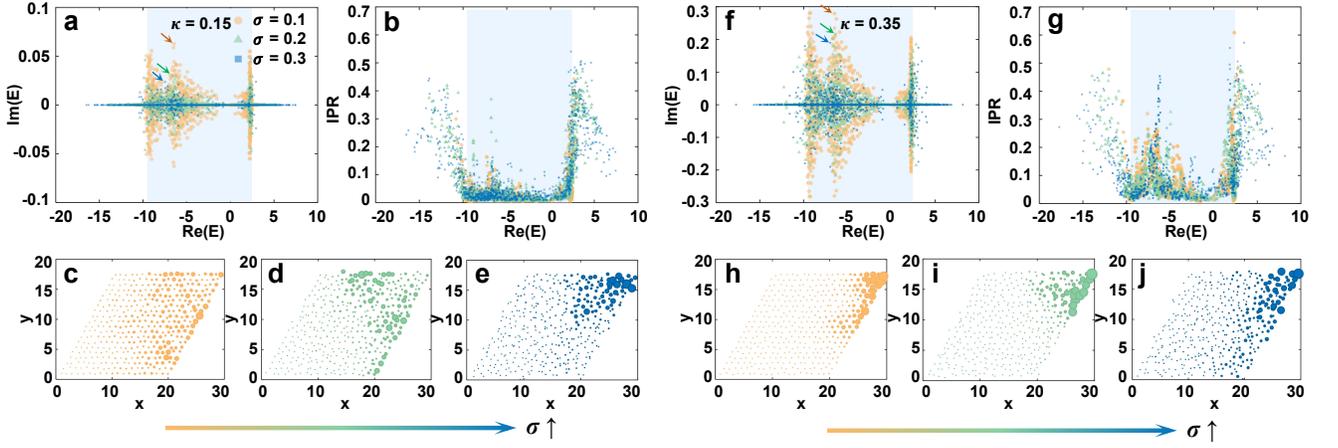

FIG. S4. The influence of structural disorder strength $\sigma$ on the energy spectrum. (a) Energy spectra of Eq. 1 at different $\sigma$ when $\kappa = 0.15, N = 20$. (b) Corresponding IPR [Eq. 2 in the main text] at different $\sigma$. (c)-(e) The state density distribution of the eigenstates pointed by arrows in (a). (f)-(j) Similar to (a)-(e), except $\kappa = 0.35$.

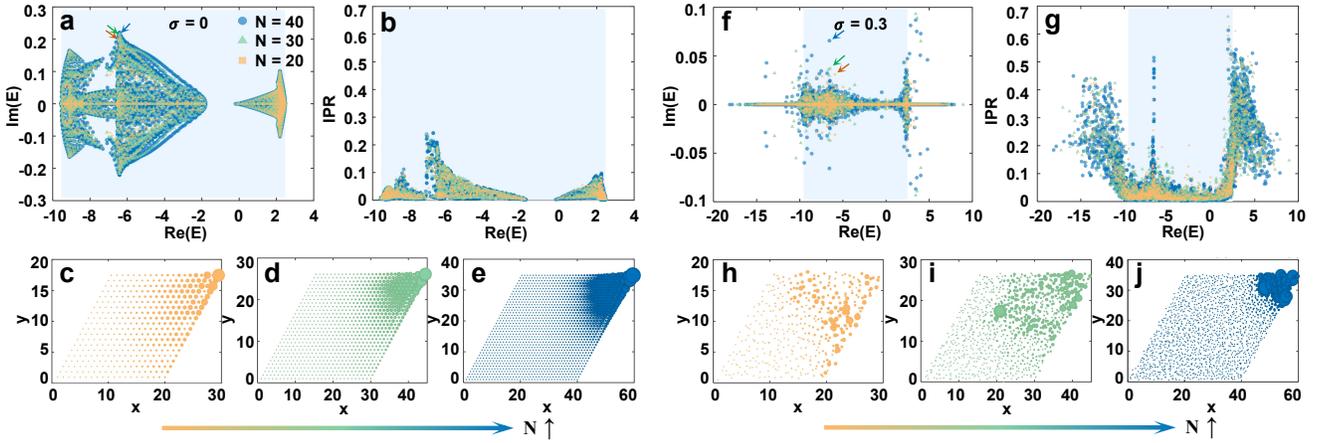

FIG. S5. The influence of system size $N$ on the energy spectrum. (a) Energy spectra of Eq. 1 at different $N$ when $\sigma = 0, \kappa = 0.15$. (b) Corresponding IPR [Eq. 2 in the main text] at different $N$. (c)-(e) The state density distribution of the eigenstates pointed by arrows in (a). (f)-(j) Similar to (a)-(e), except $\sigma = 0.3$.

## IV. COMPARISON OF THE PHYSICAL PROPERTIES IN SKIN-DOMINATED AND FULL ENERGY REGIONS

Focusing on the skin-dominated region, we also calculated the average inverse participation ratio $\langle \text{IPR} \rangle$ and the average of the square of the imaginary part $\langle \text{Im}(E)^2 \rangle$ as a function of disorder strength $\sigma$. As shown in Fig. S6(a), when $\kappa = 0.15$, as $\sigma$ increases, the non-reciprocal strength is not sufficient to resist the destruction of the non-Hermitian skin states in the clean system by disorder, and the $\langle \text{IPR} \rangle$ decreases first. As $\sigma$ continues to increase, the atomic displacement causes changes in bond angles and strengths in the system, making the eigenstates more localized. Therefore, the $\langle \text{IPR} \rangle$ increases when $\sigma$ is large. At the same time, disorder causes the number of states with $\text{Im} = 0$ to increase, and the number of states with $\text{Im} \neq 0$ to decrease, which causes the system's imaginary energy $\langle \text{Im}(E)^2 \rangle$, a new physical quantity that characterizes the imaginary part information of the system, to decrease with the increase of $\sigma$. Even though the imaginary component of the system is weakened, the states with imaginary parts that remain still show an amplification enhancement effect, as shown in Fig. 3 of the main text. When the non-reciprocal strength is increased to 0.4, the non-reciprocal strength is sufficient to resist the damage caused by disorder, and the $\langle \text{IPR} \rangle$ remains almost unchanged in the skin-dominated region. $\langle \text{Im}(E)^2 \rangle$ and the number of states with/without imaginary parts have the same trend as the former.

When we focus on the full energy region, as shown in Fig. S7(a), $\langle \max(\text{Im}(E)) \rangle$ first decreases sharply due to

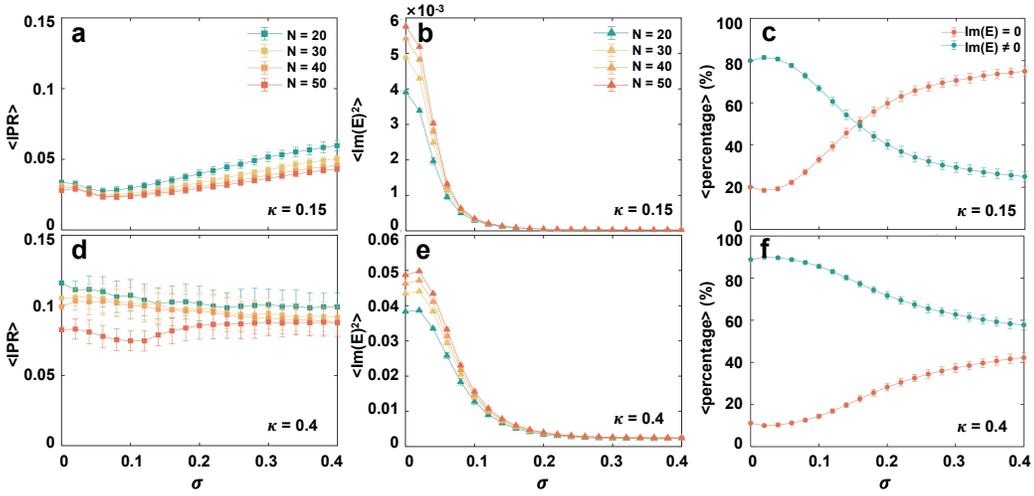

FIG. S6. Measurement of IPR and Im(E)². (a),(b) Average of IPR and Im(E)² for eigenstates at different sizes as a function of σ for κ = 0.15 in the skin-dominated region. (c) The percentage of imaginary parts as a function of σ when κ = 0.15 for N = 20. (d),(e) Average of IPR and Im(E)² for eigenstates at different sizes as a function of σ for κ = 0.4 in the skin-dominated region. (f) The percentage of imaginary parts as a function of σ when κ = 0.4 for N = 20. Each data point in the figure is the average of 400 disordered configurations.

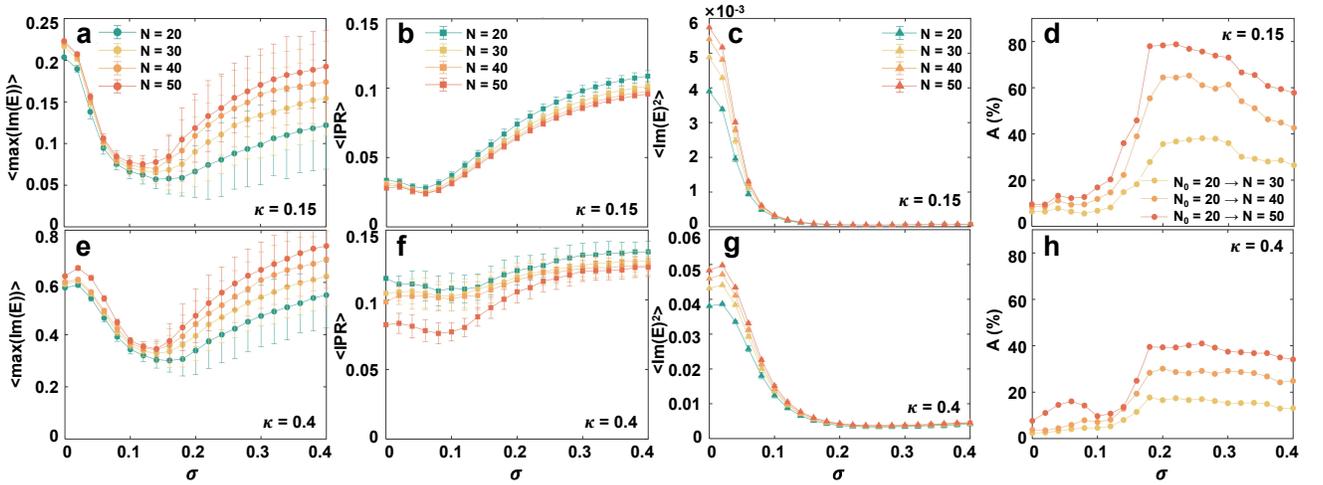

FIG. S7. Anomalous scaling of amplification rate max(Im(E)) in the full energy region. (a)-(c) Average of max(Im(E)), IPR and Im(E)² for all eigenstates at different sizes as a function of σ for κ = 0.15 in the full energy region. (e)-(g) Average of max(Im(E)), IPR and Im(E)² for all eigenstates at different sizes as a function of σ for κ = 0.4 in the full energy region. (d),(h) Based on the system size N = 20, the amplification enhancement ratio A [Eq. 3 in the main text] varies as a function of σ at different sizes, when κ = 0.15 and κ = 0.4, respectively. Each data point in the figure is the average of 400 disordered configurations.

disorder, which is independent of size. After σ is greater than 0.1, ⟨max(Im(E))⟩ shows a significant increase with increasing size. This is because more isolated states appear at the ends of the skin-dominated region, which appear localized due to Anderson localization, although they also gain an imaginary part. Similarly, Figs. S7(d) and S7(h) show the amplification enhancement ratio A after the size increases based on size N = 20. When κ = 0.15, the A reaches 80%. When κ = 0.4, the A is 40%, this is because the non-reciprocal strength (κ = 0.4) has driven the system to have a higher imaginary part, so the amplification enhancement effect is not obvious. However, it is important to note here that although a more pronounced amplification enhancement ratio has emerged in the full energy space, we should focus more on the skin-dominated region and eliminate the interference from the Anderson localized states.

Figure S8 shows the changes of ⟨max(Im(E))⟩, ⟨IPR⟩ and ⟨Im(E)²⟩ with σ under PBC. We can see that the finite size effect disappears and the change trends under different sizes are consistent. After σ increases, isolated states

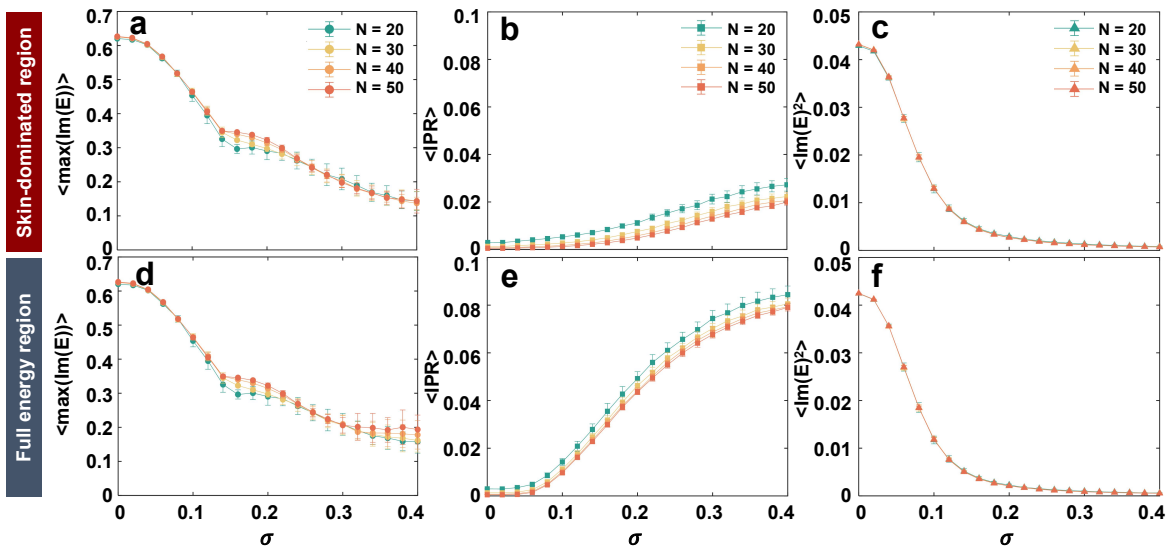

FIG. S8. The variation of energy spectrum under periodic boundary conditions (PBC). (a)-(c) Average of max(Im(E)), IPR and Im(E)$^2$ for all eigenstates of different sizes as a function of $\sigma$ for $\kappa = 0.15$ under PBC in the skin-dominated region. (d)-(f) Average of max(Im(E)), IPR and Im(E)$^2$ for all eigenstates of different sizes as a function of $\sigma$ for $\kappa = 0.15$ under PBC in the full energy region. Each data point in the figure is the average of 50 disordered configurations.

that are not affected by boundary conditions appear at the ends of the skin-dominated region. Therefore, the $\langle \text{IPR} \rangle$ in the full energy region is higher than that in the skin-dominated region. This also confirms that non-Hermitian critical scaling is *distinct* from the amplification caused by isolated impurities. Therefore, our focus is placed on the skin-dominated region.

## V. UNDERSTANDING FEEDBACK LOOPS IN REAL SPACE

Structural disorder randomly perturbs atomic positions, which leads to variation in bond angles and strength. This phenomenon can cause certain bond couplings to become stronger or more asymmetric (non-reciprocal) compared to their neighbors. Bond configurations containing net non-reciprocal closed cycles become local feedback loops that lead to amplification. Figure S9 shows several other typical local feedback loops. We can see that for some states with Im(E) $\neq 0$, a closed loop with a net circulation can be formed. These different feedback loops compete with each other to induce the emergence of non-Hermitian critical scaling. Figure S10 is the local feedback loop corresponding to Fig. S9 in real space. Because the first five points with the largest state density distribution may come from different orbits of the same atom, they have the same coordinates in real space, so some points in the local feedback loop overlap. From the real space loop, we can still get a net circulation in the darker main circulation path, which allows the wave function to be amplified or annihilated along the loop as time passes. However, for the eigenstates with Im(E) = 0, as shown in Figs. S11 and S12, the coupling with the main contribution in the local feedback loop cannot form a closed loop, which indicates that the wave function cannot be amplified or annihilated over time, but is still pumped to the corners of the system by non-reciprocal coupling.

## VI. ANALYTICAL EXPLANATION OF SKIN COMPETITION

To explain the mechanism behind the surprising observation that Im($E$) can increase due to structural disorder, we show how the essential physics can be captured by a simplified competing chain model. In general, structural disorder entails atomic displacements, altering the number and strength of couplings between atoms and introducing asymmetry (Fig. S1(c)). As a result, this system contains many coupled chains with different extents of hopping asymmetries, leading to skin competition, as shown in Fig. 4(c1) of the main text. In a 2D system (or higher), we can always find two 1D chains with significantly different skin strengths, [Fig. 4(c2) of the main text]. These two chains are coupled through many other skin chains. As an approximation, we can treat these coupling chains as Hermitian

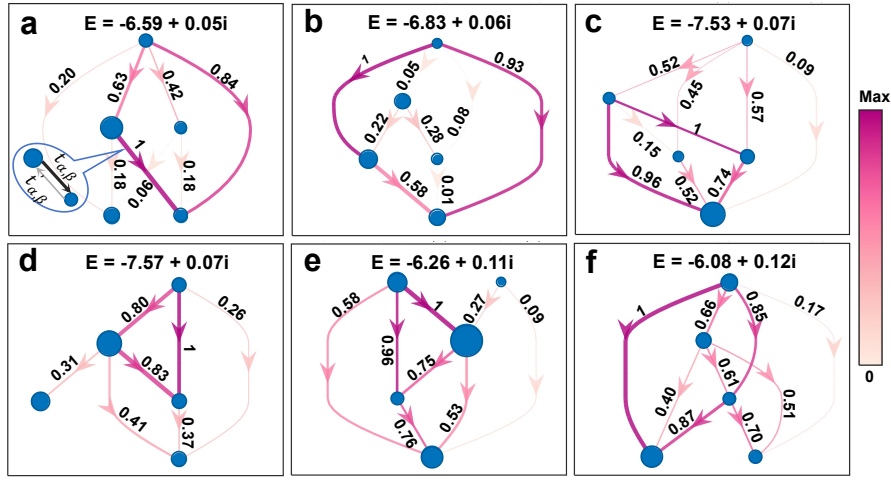

FIG. S9. **Local feedback loops.** Illustration of several other typical local feedback loops in the skin-dominated region with Im(E) $\neq$ 0. Disk size reflects the state occupancies, $|\psi_{i\alpha}|^2$. Arrows indicate net hopping directions, with color intensity representing effective hopping strength $G = \sqrt{t_{\alpha,\beta} t'_{\alpha,\beta}}$ and line width representing actual transition amplitude $W = G\sqrt{|\psi_{i\alpha}\psi_{j\beta}|}$, $\alpha$ and $\beta$ represent the orbital with the largest component of the wave function.

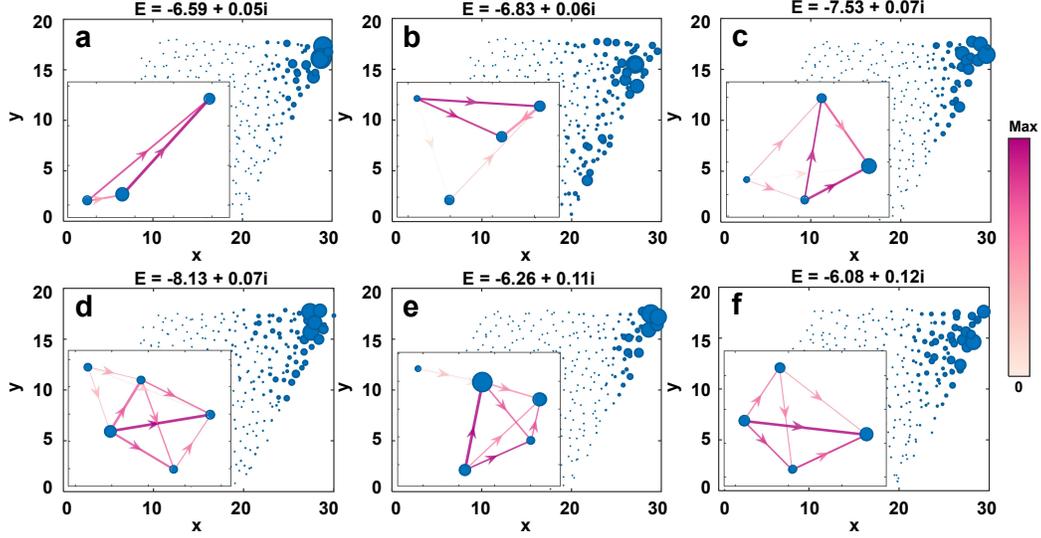

FIG. S10. **Local feedback loop and state density distribution of eigenstates with Im(E) $\neq$ 0 corresponding to Fig. S9 in real space.** Disk size reflects the state occupancies, $|\psi_{i\alpha}|^2$. Arrows indicate net hopping directions, with color intensity representing effective hopping strength $G = \sqrt{t_{\alpha,\beta} t'_{\alpha,\beta}}$ and line width representing actual transition amplitude $W = G\sqrt{|\psi_{i\alpha}\psi_{j\beta}|}$, $\alpha$ and $\beta$ represent the orbital with the largest component of the wave function.

chains, even though the results do not differ significantly without this assumption. They act as a bridge between the two 1D chains with significantly different skin strengths. Further simplifying, we can view the effect of these coupling chains as a weak coupling $\tilde{\delta}$ between the two 1D chains (Chain $A$ and chain $B$).

The simplest way to model such inter-chain interactions is to model them as a pair of Hatano-Nelson chains with weak coupling $\tilde{\delta}$, $H_A$ and $H_B$ are the respective Hamiltonians of the chains. Then, the toy model can be written as,

$$\tilde{\mathcal{H}} = H_A \oplus H_B + \sigma_x \otimes \tilde{\delta}. \tag{4}$$

Here,

$$H_\eta = \sum_x t_\eta \left[ \exp(\gamma_\eta)|x\rangle\langle x+1| + \exp(-\gamma_\eta)|x\rangle\langle x-1| \right]. \tag{5}$$

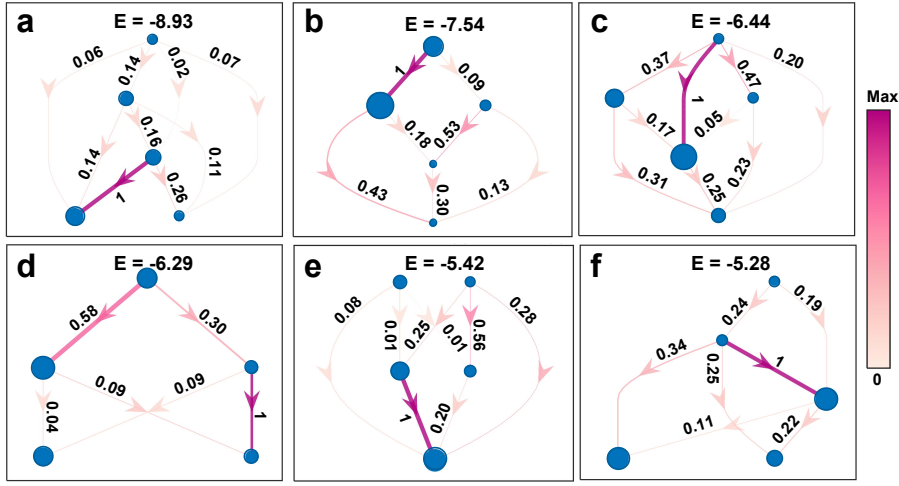

FIG. S11. Local feedback loops. Illustration of several other typical local feedback loops in the Anderson-dominated region with Im(E) = 0. Disk size reflects the state occupancies, $|\psi_{i\alpha}|^2$. Arrows indicate net hopping directions, with color intensity representing effective hopping strength $G = \sqrt{t_{\alpha,\beta} t'_{\alpha,\beta}}$ and line width representing actual transition amplitude $W = G\sqrt{|\psi_{i\alpha}\psi_{j\beta}|}$, $\alpha$ and $\beta$ represent the orbital with the largest component of the wave function.

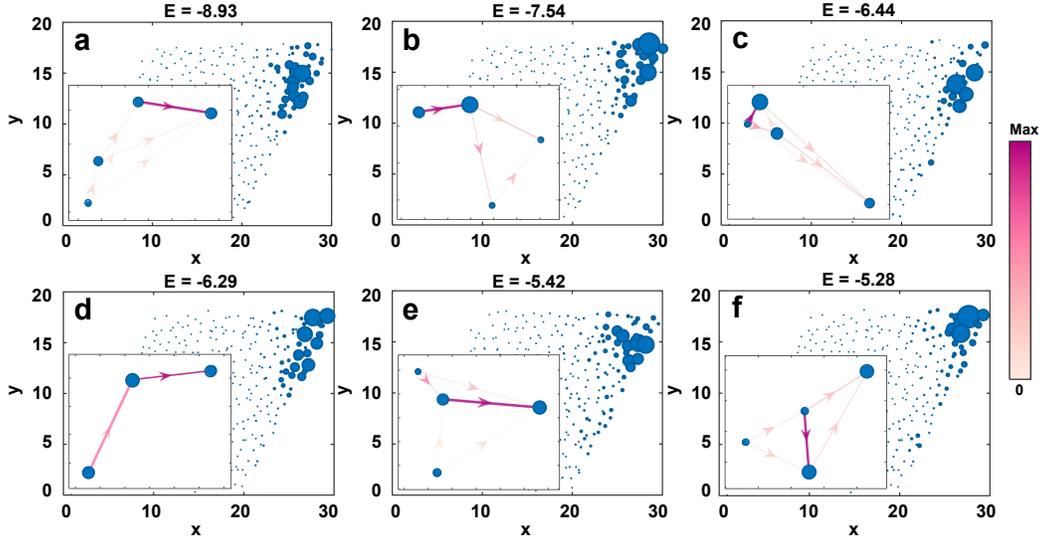

FIG. S12. Local feedback loop and state density distribution of eigenstates with Im(E) = 0 corresponding to Fig. S11 in real space. Disk size reflects the state occupancies, $|\psi_{i\alpha}|^2$. Arrows indicate net hopping directions, with color intensity representing effective hopping strength $G = \sqrt{t_{\alpha,\beta} t'_{\alpha,\beta}}$ and line width representing actual transition amplitude $W = G\sqrt{|\psi_{i\alpha}\psi_{j\beta}|}$, $\alpha$ and $\beta$ represent the orbital with the largest component of the wave function.

with $\eta = A, B$, and $t_\eta \exp(\pm\gamma_\eta)$ representing the amplitudes of nearest-neighbor hopping to the left/right. Note that $\gamma_\eta < 0$, to be consistent with the skin direction shown in the main text.

Importantly, even though we anticipate non-Hermitian skin competition, we do not need to stipulate any particular relation between $\gamma_A$ and $\gamma_B$, other than that they are different in general. To characterize their competition, we use the similarity transformation $V = \sum_x \exp((\gamma_A + \gamma_B)x/2)\mathbb{I} \otimes |x\rangle\langle x|$ where $\mathbb{I}$ is the $2 \times 2$ identity matrix corresponding to the $A, B$ subspace. This eliminates the asymmetric common part of the two coupled chains. Therefore, the Hamiltonian is modified to

$$\mathcal{H} = \frac{1}{\sqrt{t_A t_B}}V\tilde{\mathcal{H}}V^{-1} = \frac{H}{t} \oplus tH^\dagger + \sigma_x \otimes \delta, \tag{6}$$

where $\delta = \tilde{\delta}/\sqrt{t_A t_B}$ and $t = \sqrt{t_B/t_A}$, $H = \sum_x \exp(\Delta\gamma)|x\rangle\langle x+1| + \exp(-\Delta\gamma)|x\rangle\langle x-1|$, with $\Delta\gamma = (\gamma_A - \gamma_B)/2$, represent the net skin strength between two chains. Under open boundary conditions (OBCs), we can still write the Hamiltonian in the form $\mathcal{H}(k)$, where the $k$ is not real, but complex-deformed into the complex General Brillouin Zone (GBZ).

Studying $\mathcal{H}$ is more straightforward because the uncoupled chains, $H/t$ and $tH^\dagger$, naturally have opposite skin directions. Even though the chains are coupled, their weak coupling means they retain state distribution trends similar to the uncoupled case, at least to zeroth order. Therefore, the wavefunction ($\mathcal{H}\psi = E\psi$) can be

$$\psi = \sum_x \begin{pmatrix} \psi_{x,A} \\ \psi_{x,B} \end{pmatrix} \otimes |x\rangle , \quad \psi_{x,\eta} = c_{z_\eta} z_\eta^x, (\eta = A, B) \tag{7}$$

with the eigenvalue

$$\lim_{\delta \ll 1} \det(\mathcal{H} - E) = \det\left(\frac{1}{t}H(z_A) - E\right) = \det(tH^\dagger(z_B) - E) = 0, \tag{8}$$

ignoring the effect of weak coupling. And these eigenfunctions give 2 solutions of $z_{A,B}$,

$$\begin{aligned} z_{A,\pm}^x &= \mathrm{e}^{-\Delta\gamma \pm i\mathcal{K}_A}, & 2\cos(\mathcal{K}_A) &= tE, \\ z_{B,\pm}^x &= \mathrm{e}^{\Delta\gamma \pm i\mathcal{K}_B}, & 2\cos(\mathcal{K}_B) &= E/t, \end{aligned} \tag{9}$$

with complex momentum $\mathcal{K}_{A,B}$.

Since non-Hermitian skin systems are boundary-sensitive, the effect of weak coupling at the boundary cannot be ignored. The boundary constraints give

$$\begin{cases} E\psi_{A,1} &= \frac{1}{t}\mathrm{e}^{\Delta\gamma}\psi_{A,2} + \delta\psi_{B,1}, \\ E\psi_{A,L} &= \frac{1}{t}\mathrm{e}^{-\Delta\gamma}\psi_{A,L-1} + \delta\psi_{B,L}, \\ E\psi_{B,1} &= t\mathrm{e}^{-\Delta\gamma}\psi_{B,2} + \delta\psi_{A,1}, \\ E\psi_{B,L} &= t\mathrm{e}^{\Delta\gamma}\psi_{B,L-1} + \delta\psi_{A,L}, \end{cases} \xrightarrow{\text{bulk}} \begin{cases} \delta\psi_{B,1} &= \frac{1}{t}\mathrm{e}^{-\Delta\gamma}\psi_{A,0}, \\ \delta\psi_{B,L} &= \frac{1}{t}\mathrm{e}^{\Delta\gamma}\psi_{A,L+1}, \\ \delta\psi_{A,1} &= t\mathrm{e}^{\Delta\gamma}\psi_{B,0}, \\ \delta\psi_{A,L} &= t\mathrm{e}^{-\Delta\gamma}\psi_{B,L+1}, \end{cases} \tag{10}$$

with bulk equations $E\psi_{A,n} = \frac{1}{t}\mathrm{e}^{\Delta\gamma}\psi_{A,n+1} + \frac{1}{t}\mathrm{e}^{-\Delta\gamma}\psi_{A,n-1}$ and $E\psi_{B,n} = t\mathrm{e}^{-\Delta\gamma}\psi_{B,n+1} + t\mathrm{e}^{\Delta\gamma}\psi_{B,n-1}$ ($n = 1, 2, \cdots, L$).

For this particular ansatz, there is spatial inversion symmetry $\mathcal{I} = \sum_x \mathbb{I} \otimes |x\rangle\langle L+1-x|$ which transforms the left skin effect into the right skin effect, and the rotation operation $\mathcal{R} = \sum_x \sigma_x \otimes |x\rangle\langle x|$ allows us to exchange the two coupled chains. This means that the parameter region $\Delta\gamma < 0$ and $t < 1$ is sufficient to cover the entire parameter space. However, we note that such a symmetry only serves to simplify the derivations, and is by no means essential in the existence of the system size-critical behavior.

Due to the weak couplings, the distribution of the wavefunction aligns with the skin direction of the decoupled chains with $\delta = 0$ satisfying $|\psi_{A,1}| \gg |\psi_{A,L}|$ and $|\psi_{B,L}| \gg |\psi_{B,1}|$. Therefore, the left sides of the first and last lines Eq. 10 of boundary equations Eq. 10 are close to 0 and can be approximated as

$$\begin{cases} \delta\psi_{B,1} &= \frac{1}{t}\mathrm{e}^{-\Delta\gamma}\psi_{A,0}, \\ \delta\psi_{A,L} &= t\mathrm{e}^{-\Delta\gamma}\psi_{B,L+1}, \end{cases} \xrightarrow{\Delta\gamma < 0, \delta \ll 1} \begin{cases} c_A &= c_{z_{A+}} = -c_{z_{A-}}, \\ c_B &= c_{z_{B+}} z_{B+}^{L+1} = -c_{z_{B-}} z_{B-}^{L+1}. \end{cases} \tag{11}$$

And the other boundary conditions give the constraint on momenta

$$\begin{cases} \delta\psi_{B,L} &= \frac{1}{t}\mathrm{e}^{\Delta\gamma}\psi_{A,L+1}, \\ \delta\psi_{A,1} &= t\mathrm{e}^{\Delta\gamma}\psi_{B,0}, \end{cases} \to \begin{cases} -\frac{c_B t}{c_A}\delta\sin(\mathcal{K}_B)\mathrm{e}^{\Delta\gamma(L-1)} = \sin(\mathcal{K}_A(L+1)), \\ -\frac{c_A}{c_B t}\delta\sin(\mathcal{K}_A)\mathrm{e}^{\Delta\gamma(L-1)} = \sin(\mathcal{K}_B(L+1)). \end{cases} \tag{12}$$

with $z_{A\pm} = \exp(-\Delta\gamma \pm i\mathcal{K}_A)$ and $z_{B\pm} = \exp(\Delta\gamma \pm i\mathcal{K}_B)$ from Eq. 9.

If the absolute value on the left side of the equation is less than 1, then $\mathcal{K}_{A/B}$ is real (Eq. 9). Otherwise, $\mathcal{K}_{A/B}$ has non-zero imaginary part, and $\sin(\mathcal{K}_{A/B}(L+1)) \xrightarrow{L \gg 1} \exp(i\mathcal{K}_{A/B}(L+1))/2i$ with a negative imaginary part. As a result, there exists 3 distinct scenarios with characterized by the reality of the momenta $\mathcal{K}_{A/B}$:

- $\mathcal{K}_{A,B}$ are real numbers, indicating that the spectrum is purely real. In this case, the effect of weak coupling can be completely neglected, and the spectrum coincides with that of the decoupled model. This generally occurs in small-size systems with weak coupling, and the corresponding energy range $[-2t, 2t]$.

- $\mathcal{K}_{A,B}$ have exactly one real number, the spectrum is also real, it only exists in the energy range $[-2/t, -2t]\,\&\,[2t, 2/t]$.

- $\mathcal{K}_{A,B}$ are both complex numbers, and the spectrum becomes complex. The approximate result

$$\lim_{L \gg 1} \sin(\mathcal{K}_{A,B}(L+1)) \xrightarrow{\mathrm{Im}\mathcal{K}_{A,B}<0} \exp(i\mathcal{K}_{A,B}(L+1))/(2i)$$

comes in useful. Taking the $1/(L+1)$-th power on both sides of the equation,

$$\begin{cases} i\mathcal{K}_A = \Delta\gamma + \frac{1}{L+1}\left(\log\left(-2i\delta\mathrm{e}^{-2\Delta\gamma}\right) + \log\sin\mathcal{K}_B + \log\left(\frac{c_B t}{c_A}\right)\right) + iq_A \\ i\mathcal{K}_B = \Delta\gamma + \frac{1}{L+1}\left(\log\left(-2i\delta\mathrm{e}^{-2\Delta\gamma}\right) + \log\sin\mathcal{K}_A - \log\left(\frac{c_B t}{c_A}\right)\right) + iq_B, \end{cases} \tag{13}$$

with real momentum $q_A, q_B$. And the bulk equations $2\cos\mathcal{K}_A/t = 2t\cos\mathcal{K}_B = E$ give the other constraint

$$\tan\frac{\mathcal{K}_A + \mathcal{K}_B}{2}\tan\frac{\mathcal{K}_A - \mathcal{K}_B}{2} = \frac{1-t^2}{1+t^2}. \tag{14}$$

In general, since the absolute value of $z$ depends on the specific energy, the most noticeable change in GBZ occurs at the energy with the largest imaginary part, which also controls the dominant dynamical evolution. Therefore, we now focus on the case where the energy has the maximum imaginary part and zero real part. This energy is unique, and the real momenta $K_A$ and $K_B$ satisfy $\mathrm{Re}(\mathcal{K}_A) = \mathrm{Re}(\mathcal{K}_B) = \pi/2$, i.e., $(q_A + q_B)/2 = \pi/2$. That is, the GBZ is given by

$$\log|z_{A\pm}| = -\Delta\gamma \pm \left(\mathrm{Im}(F) + \mathrm{Im}\left(\mathrm{atan}\left(\frac{1-t^2}{1+t^2}\cot(F)\right)\right)\right) + iq_1,$$

$$\log|z_{B\pm}| = \Delta\gamma \pm \left(\mathrm{Im}(F) - \mathrm{Im}\left(\mathrm{atan}\left(\frac{1-t^2}{1+t^2}\cot(F)\right)\right)\right) + iq_2, \tag{15}$$

where real $q_1, q_2 \in [-\pi, \pi]$ and the approximation $\frac{1}{L+1}\log\sin\mathcal{K}_{A/B} = 0$ s applied on the right-hand side of Eq. 13,

$$iF = i\pi/2 + \Delta\gamma + \frac{1}{L+1}\log\left(-2i\delta\mathrm{e}^{-2\Delta\gamma}\right). \tag{16}$$

Simply put, if we take the absolute value of the GBZ corresponding to the energy with the largest imaginary part as the representative value of most physical significance, we can express the GBZ as in Eq. 15, but with the real momentum ranging from $[-\pi, \pi]$.

Hence, combining the above scenarios, the complex piece of the spectrum is given by

$$E_{\mathrm{cx}} \in \left\{2t\cos(\mathcal{K}_B)\,\Big|\quad \mathcal{K}_B = k + i\mathrm{Im}(F) - i\mathrm{Im}\left(\mathrm{atan}\left(\frac{1-t^2}{1+t^2}\cot(F)\right) + i\pi/2\right)\right\}, \tag{17}$$

with real $k \in [-\pi, \pi]$ and $F$ from Eq. 16. The periodicity of $k$ means that $E_{\mathrm{cx}}$ forms a loop in the complex plane. The real energy piece, if it exists, is a segment given by

$$E_{\mathrm{re}} \in E_{\mathrm{real}}\,\&\,-E_{\mathrm{real}}, \quad E_{\mathrm{real}} = [\max(\mathrm{Re}(E_{\mathrm{cx}})), 2/t]. \tag{18}$$

if $\max(\mathrm{Re}(E_{\mathrm{cx}})) < 2/t$. In all, the system energy spectrum is given by

$$E_{\mathrm{eff}} \in E_{\mathrm{re}} \cup E_{\mathrm{cx}}. \tag{19}$$

Note that $E_{\mathrm{re}}$ is an empty set if $\max(\mathrm{Re}(E_{\mathrm{cx}})) > 2/t$, as shown in Fig. S13(a). The analytical results (blue line) are consistent with the numerical results (black dots).

Focusing on the spectrum's largest imaginary part, with the corresponding energy denoted as $E'$, we have

$$E' = -2it\sinh\left(\mathrm{Im}(F) - \mathrm{Im}\left(\mathrm{atan}\left(\frac{1-t^2}{1+t^2}\cot(F)\right)\right)\right). \tag{20}$$

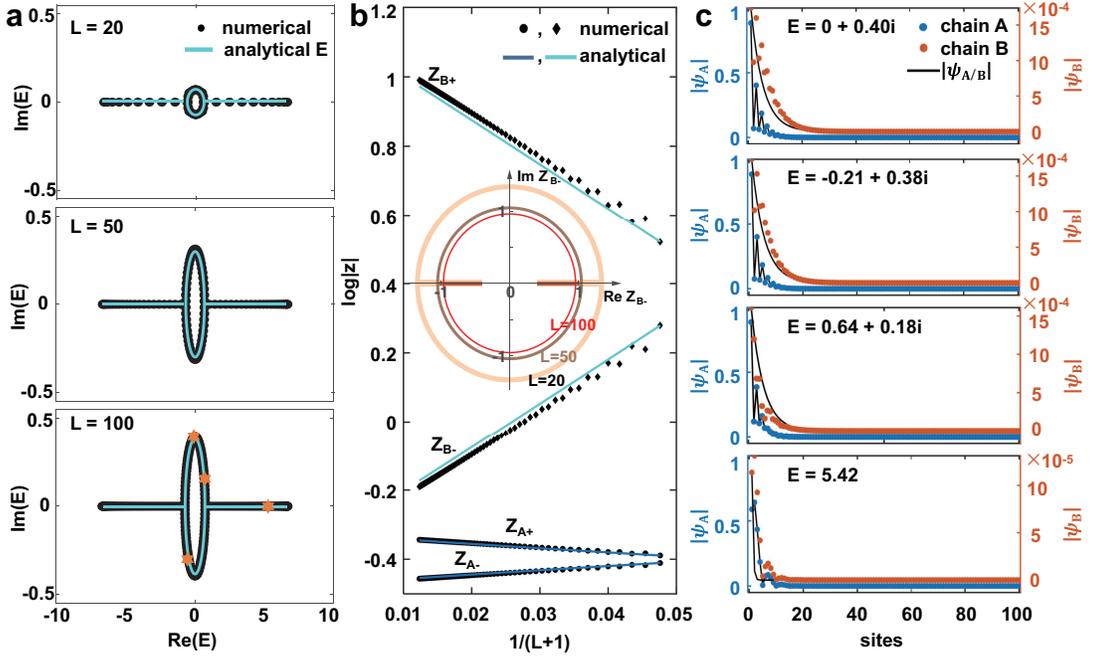

FIG. S13. (a) Energy spectrum of model Eq. 6 with different size: Numerical vs. Analytical results [Eq. 19]. (b) Agreement of the size-dependent inverse decay length $\log|z|$ obtained in two different ways: numerically, by solving $\det(\mathcal{H}(z) - E') = 0$ from the energy with the largest imaginary part in the numerical eigenspectrum; analytically, via the size-dependent GBZ we derived [Eq. 15]. (c) Numerical (blue and orange dots) and analytical (black lines) results of the illustrative eigenfunctions at different energies when $L = 100$, marked with orange stars in (a). The parameters are $t = 0.3$, $\Delta\gamma = 0.4$, $\delta = 10^{-3}$.

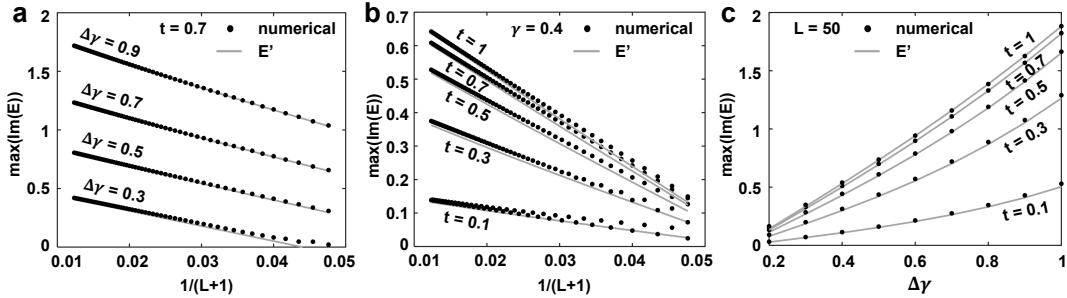

FIG. S14. The critical scaling behavior of the spectrum. The dependence of the maximum imaginary part of energy $\max(\mathrm{Im}(E))$ on parameters $\Delta\gamma$, $t$, and $L$ is compared between numerical and analytical results, $E'$ [Eq. 21]. (a) As $\Delta\gamma$ increases, the size effect of $E'$ shows that stronger non-Hermitian effects lead to a larger imaginary part, like the PBC case; (b) As $t$ varies, the size effect of $E'$ shows that the smaller the difference between the strengths of the two coupled chains, the larger the imaginary part, with the maximum at $t = 1$; (c) The variation of $E'$ with $\Delta\gamma$ and $t$ also shows that the largest imaginary part occurs at $t = 1$ and larger $\Delta\gamma$. Parameter $\delta = 10^{-3}$.

The numerical and analytical results of GBZ with maximum imaginary energy (Fig. S13(b)) show that as size increases, the size-dependent GBZ analytical solution, the inverted decay length $\log|z|$ by solving Eq. 15 is consistent with the numerical solution by solving $\det(\mathcal{H}(z) - E') = 0$ from the energy with the largest imaginary part in the numerical eigenspectrum. We can intuitively see that $\log|z|$ maintains a linear relationship with $1/(L+1)$. In addition, the eigenfunctions at different energies when $L = 100$ (Fig. S13(c)), marked with orange stars in Fig. S13(a), and their numerical results fit well with the analytical results.

Additionally, the dependence of the maximum imaginary part of energy on the parameters $\Delta\gamma$, $t$, and $L$ (Fig. S14(a)) shows that the numerical solution and the analytical solution fit well. As the size increases, the maximum imaginary part shows obvious size dependence and shows an increasing behavior, that is, the non-Hermitian critical

scaling. When we fix $\Delta\gamma$, in Fig. S14(b), $t$ represents the quadratic ratio of the coupling strengths of the two chains. When $t = 1$, $t_A = t_B$, max(Im(E)) depends on the size linearly, and has the same behavior even when $t$ is very small. In Fig. S14(c), with a fixed size $L = 50$, as $\Delta\gamma$ increases (the net skin strength between the two chains), the max(Im(E)) also increases, and the system shows a strong NHSE. Since we hide the disorder in $\Delta\gamma$, this also indicates that for a 2D disordered system, the energy spectrum changes regularly with the increase of $\sigma$. Therefore, this also proves the stochasticity-induced non-Hermitian skin criticality mentioned in the main text.

Going back to the Hamiltonian before the similarity transformation (Eq. 4), the effective energy can be written as:

$$\tilde{E}_{\text{eff}} = \sqrt{t_A t_B} \times E_{\text{eff}}, \tag{21}$$

with $\delta = \tilde{\delta}/\sqrt{t_A t_B}$, $t = \sqrt{t_B/t_A}$, and $\Delta\gamma = (\gamma_A - \gamma_B)/2$ and $E_{\text{eff}}$ from Eq. 19 . Obviously, for the general model, the energy spectra show size dependence.

---

\* phyhuij@nus.edu.sg
† zhanglei@sxu.edu.cn
‡ yeesin_ang@sutd.edu.sg
§ phylch@nus.edu.sg


\end{document}